\documentclass[reprint,prd,amsmath,amssymb,aps,floatfix,twocolumn,nofootinbib]{revtex4-2}

\usepackage[english]{babel}
\usepackage{graphicx}
\usepackage{amssymb,amsmath}
\usepackage{xcolor}
\usepackage{aas_macros}
\usepackage{siunitx}
\usepackage[caption=false]{subfig}
\usepackage{booktabs}
\usepackage{multirow}
\usepackage{float}
\usepackage{url}
\usepackage[colorlinks=true, allcolors=blue]{hyperref}

\begin{document}

\title{MeV, GeV and TeV neutrinos from binary-driven hypernovae}

\author{S. Campion,$^{1,2}$ J. D. Uribe-Suárez,$^{3,4}$ J.~D.~Melon Fuksman,$^{5}$ J. A. Rueda,$^{3,6,7,8,9}$ 
} 

\affiliation{$^{1}$Dipartimento di Fisica, Sapienza Universit\`a di Roma, P. le Aldo Moro 2, I-00185 Rome, Italy}
\affiliation{$^{2}$Istituto Nazionale di Fisica Nucleare, Sezione di Roma, P. le Aldo Moro 2, I-00185 Rome, Italy}
\affiliation{$^{3}$ICRANet, Piazza della Repubblica 10, I-65122 Pescara, Italy}
\affiliation{$^{4}$Facultad de Ciencias Básicas, Universidad Santiago de Cali, Campus Pampalinda, Calle 5 No. 6200, 760035 Santiago de Cali, Colombia}
\affiliation{$^{5}$Max Planck Institute for Astronomy, Königstuhl 17, D-69117 Heidelberg, Germany}
\affiliation{$^{6}$ICRA, Dipartimento di Fisica, Sapienza Universit\`a  di Roma, P.le Aldo Moro 5, I-00185 Roma, Italy}

\affiliation{$^{7}$ICRANet-Ferrara, Dipartimento di Fisica e Scienze della Terra, Universit\`a degli Studi di Ferrara, Via Saragat 1, I-44122 Ferrara, Italy}
\affiliation{$^{8}$Dipartimento di Fisica e Scienze della Terra, Universit\`a degli Studi di Ferrara, Via Saragat 1, I-44122 Ferrara, Italy}
\affiliation{$^{9}$INAF, Istituto di Astrofisica e Planetologia Spaziali, Via Fosso del Cavaliere 100, I-00133 Rome, Italy}

\email{stefano.campion@uniroma1.it}
\email{juan.uribe@icranet.org}
\email{fuksman@mpia.de}
\email{jorge.rueda@icra.it}

\date{\today / Received date /
Accepted date}

\begin{abstract}
We analyze neutrino emission channels in energetic ($\gtrsim 10^{52}$~erg) long gamma-ray bursts within the binary-driven hypernova model. The binary-driven hypernova progenitor is a binary system composed of a carbon-oxygen star and a neutron star (NS) companion. The gravitational collapse leads to a type Ic supernova (SN) explosion and triggers an accretion process onto the NS. For orbital periods of a few minutes, the NS reaches the critical mass and forms a black hole (BH). Two physical situations produce MeV neutrinos. First, during the accretion, the NS surface emits neutrino-antineutrino pairs by thermal production. We calculate the properties of such a neutrino emission, including flavor evolution. Second, if the angular momentum of the SN ejecta is high enough, an accretion disk might form around the BH. The disk's high density and temperature are ideal for MeV-neutrino production. We estimate the flavor evolution of electron and non-electron neutrinos and find that neutrino oscillation inside the disk leads to flavor equipartition. This effect reduces (compared to assuming frozen flavor content) the energy deposition rate of neutrino-antineutrino annihilation into electron-positron ($e^+e^-$) pairs in the BH vicinity. We then analyze the production of GeV-TeV neutrinos around the newborn black hole. The magnetic field surrounding the BH interacts with the BH gravitomagnetic field producing an electric field that leads to spontaneous $e^+e^-$ pairs by vacuum breakdown. The $e^+e^-$ plasma self-accelerates due to its internal pressure and engulfs protons during the expansion. The hadronic interaction of the protons in the expanding plasma with the ambient protons leads to neutrino emission via the decay chain of $\pi$-meson and $\mu$-lepton, around and far from the black hole, along different directions. These neutrinos have energies in the GeV-TeV regime, and we calculate their spectrum and luminosity. We also outline the detection probability by some current and future neutrino detectors.
\end{abstract}


\maketitle

%%%%%%%%%%%%%%%%%%%%%%%%%%%%%%%%%%%%%%%%%%%%%%%%%%
%%%%%%%%%%%%%%%%% BODY OF PAPER %%%%%%%%%%%%%%%%%%

%%%%%%%%%%%%%%%%%%%%%%%%%%%%%%%%%%%%%%%%%%%%%%%%%%%%
%%%%%%%%%%%%%%%%%%%%%%%%%%%%%%%%%%%%%%%%%%%%%%%%%%%%
\section{Introduction}\label{sec:1}
%%%%%%%%%%%%%%%%%%%%%%%%%%%%%%%%%%%%%%%%%%%%%%%%%%%%
%%%%%%%%%%%%%%%%%%%%%%%%%%%%%%%%%%%%%%%%%%%%%%%%%%%%

Multi-messenger astronomy is fundamental to acquiring information about the physical processes, dynamics, evolution, and structure behind the cosmic sources and unveiling their nature~\cite{2018Sci...361..147I}. With the advent of new observational facilities generating high-quality data from energetic sources, such as supernovae (SNe), gamma-ray bursts (GRBs), and active galactic nuclei (AGNs), the analysis of the multi-messenger emission becomes a necessity. Here, we aim to study the neutrino messenger
with energies from MeV to GeV-TeV for long GRBs within the binary-driven hypernova (BdHN) model. In Section \ref{sec:2}, we summarize the BdHN model and the relevant features for the present article.

The study of neutrino emission in GRBs started with the pioneering work of \citet{waxman1997high}. They examined the production of energetic neutrinos ($E_{\nu}\sim 10^{14}$~eV) arising from the photomeson production process by the interaction between very-high-energy protons ($E_p\lesssim 10^{20}$ eV) and photons emitted through synchrotron/inverse Compton (IC) radiation by accelerated electrons. Other works followed studying the neutrino production in the \textit{fireball} model of GRBs (see, e.g., \cite{waxman1998high, waxman2000neutrino, bahcall20005, meszaros2000multi}). These works show that the internal shock in the fireball produces neutrinos from pion and muon decay, mainly by two dominant processes: 1) the photomeson production ($p+\gamma\rightarrow \pi, \mu\rightarrow \nu_{\mu,e}$) that leads to $\sim 10^{14}$~eV~\cite{waxman1998high} or $\sim 10^{18}$~eV neutrinos when $\sim 10^{20}$~eV protons interact with (a few) eV photons~\cite{waxman2000neutrino}, and 2) the pion and muon production by the interaction of accelerated protons and coasting neutrons in the expanding fireball, leading to $5$--$10$ GeV neutrinos~\cite{bahcall20005,meszaros2000multi}.

There is also GRB literature on neutrinos from neutron decay, stellar collapse, or compact-object mergers can be also found in the GRB literature (see Ref.~\cite{meszaros2000multi}, and references therein). Those mechanisms are less efficient than the previous channels and produce neutrinos of energies $10\leq E_{\nu}\leq 100$~MeV, which would be difficult to detect due to the low values of $\nu N$ cross-section for low $E_{\nu}$.

Having recalled some of the classic studies of neutrinos in GRBs, we turn to the specific topic of this work. We here study neutrino production channels in the BdHN scenario, an alternative to the traditional GRB models.  The BdHN model proposes that the GRB originates in a binary composed of a carbon-oxygen (CO) star and a neutron star (NS) companion~\cite{2012ApJ...758L...7R}. The gravitational collapse of the iron core of the CO star leads to a newborn NS (hereafter $\nu$NS) and a type Ic supernova (SN) explosion. The latter triggers an accretion process onto the $\nu$NS and the NS. For orbital periods of a few minutes, the NS companion reaches the critical mass and forms a BH~\cite{2014ApJ...793L..36F, 2016ApJ...833..107B}. Systems leading to BH formation are called BdHN of type I (hereafter BdHN I). Less compact binaries do not form BHs and are BdHN II and BdHN III. In the former, the orbital period is tens of minutes. In the latter, the orbital size is even longer, marginalizing the role of the NS companion, and the system behaves as a single collapsing star leading to an SN Ic. A recent account of the main features of the BdHN subclasses, its physical phenomena, and related GRB observables from the radio to the optical, to the X-rays, to the high-energy gamma-rays, including analyses of specific GRBs can be found in Section \ref{sec:2} and Refs. \cite{2022ApJ...939...62R, 2022PhRvD.106h3004R, 2022PhRvD.106h3002B, 2022EPJC...82..778R, 2022ApJ...936..190W, 2022ApJ...929...56R}.

This article studies neutrino emission in the most energetic subclass, the BdHN I. We address two energy regimes that correspond to different mechanisms of neutrino production and times of occurrence in the BdHN I leading to the GRB event. In Section \ref{sec:3}, we analyze two processes leading to MeV neutrinos in BdHN I, i.e., the hypercritical accretion process onto the $\nu$NS and the NS companion, and onto the newborn BH, after the gravitational collapse of the NS. We investigate the role of neutrino flavor oscillations in detail. In Section \ref{sec:4}, we turn to the GeV-TeV neutrinos produced by $pp$ interactions from protons swept by the expanding $e^+e^-$ plasma created (via vacuum polarization) around the newborn BH, with protons of the medium in the vicinity as well as far from the BH site. Finally, we draw in Section~\ref{sec:5} the conclusions of our results.

%%%%%%%%%%%%%%%%%%%%%%%%%%%%%%%%%%%%%%%%%%%%%%%
%%%%%%%%%%%%%%%%%%%%%%%%%%%%%%%%%%%%%%%%%%%%%%%
\section{The BdHN model}
\label{sec:2}
%%%%%%%%%%%%%%%%%%%%%%%%%%%%%%%%%%%%%%%%%%%%%%%
%%%%%%%%%%%%%%%%%%%%%%%%%%%%%%%%%%%%%%%%%%%%%%%

Before entering into details of the BdHN model, we recall aspects of the traditional GRB models and limitations from which follow the necessity for alternative scenarios. From the progenitor viewpoint, the traditional GRB model follows the concept of \textit{collapsar}, a single massive star that collapses forming a BH and an accretion disk \citep{1993ApJ...405..273W}. It is expected that such a system generates an electron-positron ($e^-e^+$)-photon-baryon plasma, a \textit{fireball}, whose transparency leads to the GRB prompt emission \cite{1978MNRAS.183..359C, 1986ApJ...308L..43P, 1986ApJ...308L..47G, 1991ApJ...379L..17N, 1992ApJ...395L..83N}. The fireball expands as a collimated jet that reaches transparency with an ultra-relativistic Lorentz factor $\Gamma \sim 10^2$--$10^3$ \cite{1990ApJ...365L..55S, 1992MNRAS.258P..41R, 1993MNRAS.263..861P, 1993ApJ...415..181M, 1994ApJ...424L.131M}. Internal and external shocks lead to the GRB prompt and multiwavelength afterglow, e.g., by synchrotron self-Compton emission \cite{2002ARA26A..40..137M, 2004RvMP...76.1143P, 2019Natur.575..455M, 2019Natur.575..448Z}. A comprehensive, recent review of the traditional GRB model can be found in \cite{2018pgrb.book.....Z}.

A major observational constraint for GRB models arises from the association of long GRBs with type Ic SNe, discovered by the optical follow-up of the afterglow, first evidenced with the GRB 980425-SN 1998bw association \cite{1998Natur.395..670G}. Since then, many more GRB-SN associations have been confirmed \cite{2006ARA&A..44..507W, 2011IJMPD..20.1745D, 2012grb..book..169H}. Theoretically, the gravitational collapse of a single massive star would hardly lead to a collapsar, a fireball with jetted emission, and a SN explosion. There are also observational facts supporting this view. Long GRBs and SNe have widely different energetics, the latter $10^{49}$--$10^{51}$~erg, the former $10^{49}$--$10^{54}$~erg. The GRB energetics point to stellar-mass BH formation, while SNe should leave an NS as central remnant. The latter is also supported by observations of pre-SN stars which point to zero-age main-sequence (ZAMS) progenitors of $\lesssim 18~M_\odot$ \cite{2009ARA&A..47...63S, 2015PASA...32...16S}, while most theoretical models predict direct BH formation only for ZAMS masses $\gtrsim 25 M_\odot$ (see, e.g., \cite{2003ApJ...591..288H}). Therefore, it seems unlikely that the GRB and the SN originate from the very-same single-star progenitor. The GRB-SN association provides an additional observational clue for GRB models, i.e., the associated SN is of type Ic, so they are absent of hydrogen (H) and helium (He). The theoretical consensus is that SNe Ic progenitors lose their hydrogen and helium envelopes during the stellar evolution, ending as He, CO, or Wolf-Rayet (WR) stars \cite{2011MNRAS.415..773S, 2020MNRAS.492.4369T}. The preferred channel to form stripped-envelope He/CO/WR stars leading to SNe Ic are short-period binaries with a compact-star companion (e.g., an NS) that evolve through mass-transfer and common-envelope phases \cite{1988PhR...163...13N, 1994ApJ...437L.115I, 2007PASP..119.1211F, 2010ApJ...725..940Y, 2011MNRAS.415..773S, 2015PASA...32...15Y, 2015ApJ...809..131K}. From the above incomplete but representative list of theoretical and observational constraints, it seems natural to examine long GRB progenitors based on binary systems, as they can appear from binary stellar evolution channels (see, e.g., \cite{1999ApJ...526..152F, 2015PhRvL.115w1102F}). In this article, we explore the neutrino emission within the BdHN scenario of long GRBs, which we recall below.

In the BdHN model, long GRBs are produced in CO-NS binaries~\cite{2012ApJ...758L...7R, 2014ApJ...793L..36F}. These binaries might be a subclass of the ultra-stripped binaries leading to type Ic SNe (see, e.g., \cite{2013ApJ...778L..23T, 2015MNRAS.451.2123T}). Before the BdHN event leading to the GRB, they follow an evolutionary path including a first SN explosion, common-envelope phases, tidal interactions, and mass loss (see \cite{2015PhRvL.115w1102F}, and references therein for details). The second SN event, i.e., in the core collapse of the CO star, triggers the GRB. We now summarize some salient features of the BdHN model from the GRB electromagnetic emission viewpoint. Recent reviews of the BdHN model can be found in \cite{2021IJMPD..3030007R, rueda2021update, 2019Univ....5..110R}. 

The core-collapse SN and the hypercritical accretion onto the $\nu$NS and the NS companion lead to electromagnetic precursors to the prompt gamma-ray emission \cite{2016ApJ...833..107B}. We refer to \cite{2019ApJ...874...39W, 2022PhRvD.106h3002B} for numerical simulations of the accretion process onto both NSs and the associated emission. 

In BdHN I, the accretion leads the NS companion to the critical mass and forms a rotating (Kerr) BH surrounded by a magnetic field and ionized low-density matter. This \textit{tryptic} has been called the \textit{inner engine} of the high-energy emission of long GRBs \cite{2019ApJ...886...82R,rueda2020blackholic, 2021A&A...649A..75M, 2021MNRAS.504.5301R, 2022ApJ...929...56R}. The electric field induced by the magnetic field-BH gravitomagnetic interaction is initially larger than the critical field for vacuum breakdown, $E_c = m_e^2 c^3/e \hbar$. Hence, it rapidly produces an $e^+e^-$ pair plasma around the BH \cite{2021PhRvD.104f3043M, 2022EPJC...82..778R}. Such a plasma self-accelerates to ultrarelativistic velocities and engulfs baryons from the surroundings during its expansion. While it expands, the plasma swept baryons in a number depending on the direction of expansion because of the asymmetry of the matter distribution around the newborn BH (see scheme in Fig. \ref{fig:Interactionsscheme1} below, and the three-dimensional simulations in \cite{2019ApJ...871...14B, 2022PhRvD.106h3002B}).

\begin{figure*}
    \includegraphics[width=0.48\hsize,clip]{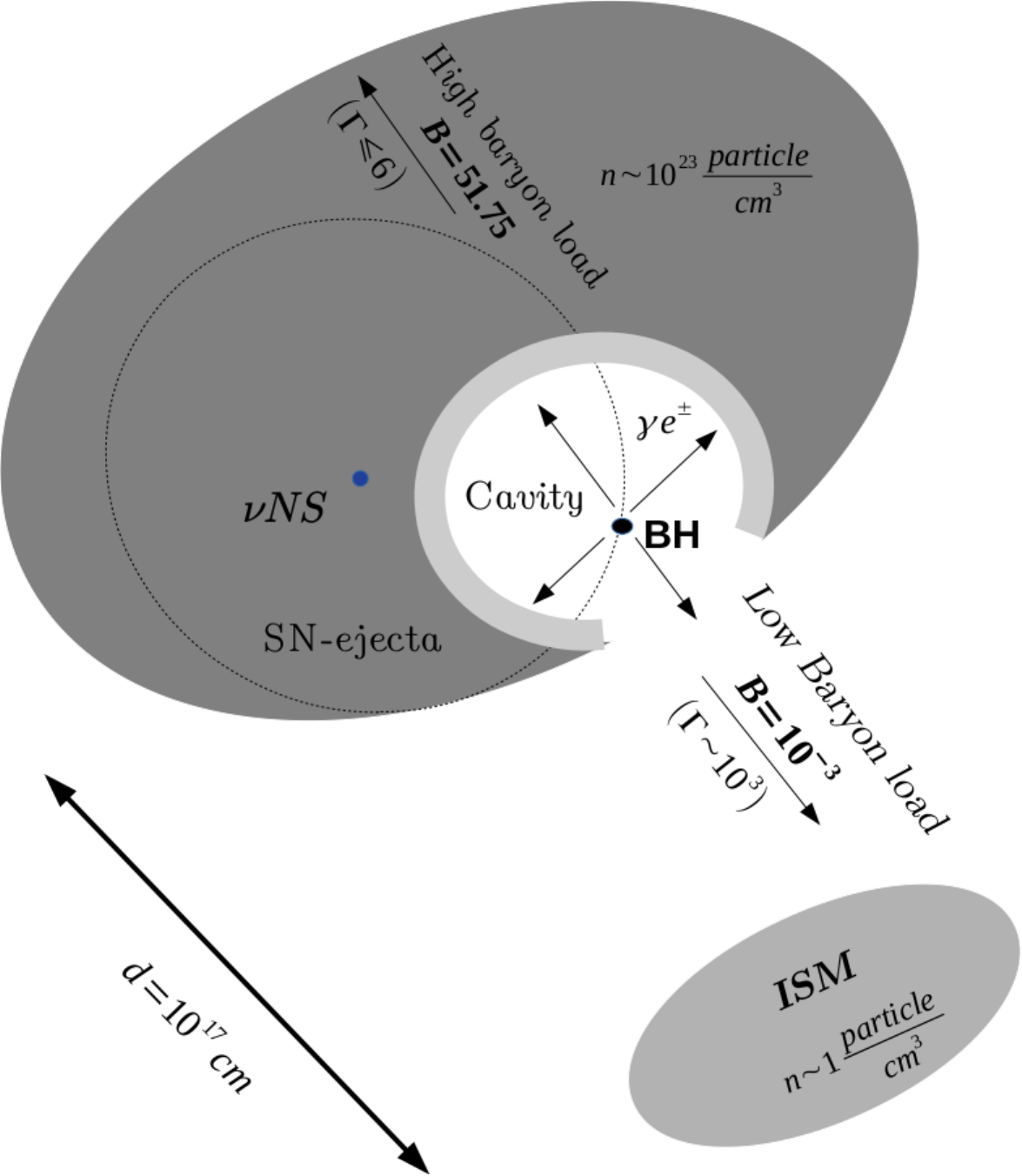}
    \caption{Scheme of the $pp$ interactions in a BdHN I. 1) The $e^+e^-$ plasma propagates inside the ejecta along directions with high baryon load, e.g., in our simulations,~$B=51.75$, reaching a Lorentz factor $\Gamma \lesssim 6$. The engulfed protons have such $\Gamma$ (see Section~\ref{S:PhysicalQuantities} for details) and interact with the protons at rest, ahead of the plasma front, and deposited all of their energy. 2) Protons engulfed by the $e^+ e^-p$ plasma propagate in the direction where the cavity is open. This plasma is loaded with a relatively low baryon content (e.g.,~$B\sim 10^{-3}$), so the plasma reaches a high Lorentz factor at transparency, $\Gamma\sim 10^2$--$10^3$. The engulfed protons have such $\Gamma$ factor and interact with the ISM protons at rest. The gray-color scale of the shaded regions highlights the different densities of the target regions in the simulation presented in this work. The dotted circular line represents the $\nu$NS-BH binary orbit, assuming the system remains bound after the SN explosion \cite{2015PhRvL.115w1102F}.}
    \label{fig:Interactionsscheme1}
\end{figure*}

Therefore, the plasma becomes transparent at different times and values of the Lorentz factor ($\Gamma$) depending on the direction of expansion. In the ultralow density region around the BH, the transparency of the plasma is reached with $\Gamma \sim 10^2$--$10^3$, leading to the ultrarelativistic prompt emission (UPE) phase (see \cite{2021PhRvD.104f3043M, 2022EPJC...82..778R}, for details). The plasma transparency along directions of higher density leads to the hard and soft X-ray flares (HXFs and SXFs) observed in the early afterglow \cite{2018ApJ...852...53R}. 

The electric field accelerates electrons from the matter surrounding the BH. Off-polar axis, electrons have non-zero pitch angles leading to synchrotron radiation losses that explain the GeV emission observed in some energetic long GRBs \cite{2019ApJ...886...82R, rueda2020blackholic, 2021A&A...649A..75M, 2022ApJ...929...56R}.

The synchrotron radiation by relativistic electrons in the SN ejecta explains the multi-wavelength (X, optical, radio) afterglow emission \cite{2018ApJ...869..101R, 2019ApJ...874...39W, 2020ApJ...893..148R}. The ejecta expands through the magnetized medium proportioned by the $\nu$NS, which, in addition, injects rotational energy into the ejecta. We refer to \cite{2022ApJ...939...62R, 2022ApJ...936..190W} for a recent analytic theoretical treatment of the above afterglow model. The synchrotron afterglow in this scenario depends only on the $\nu$NS and the SN ejecta, so it is present in all BdHN types, i.e., BdHN I, II, and III.

Finally, at about $10^6$ s after the GRB trigger, we have the optical emission from the SN ejecta due to the nuclear decay of nickel.

%%%%%%%%%%%%%%%%%%%%%%%%%%%%%%%%%%%%%%%%%%%%%%%
%%%%%%%%%%%%%%%%%%%%%%%%%%%%%%%%%%%%%%%%%%%%%%%
\section{MeV neutrinos from BdHN I}
\label{sec:3}
%%%%%%%%%%%%%%%%%%%%%%%%%%%%%%%%%%%%%%%%%%%%%%%
%%%%%%%%%%%%%%%%%%%%%%%%%%%%%%%%%%%%%%%%%%%%%%%

We devote this section to reviewing recent results on the flavor oscillations~\cite{deSalas:2017kay} in MeV-neutrinos produced in BdHNe during the accretion (onto the $\nu$NS and the NS companion) of material expelled in the SN explosion. As we shall show below, the high density of neutrinos and matter on top of the surface of the accreting NS leads to flavor oscillations and new neutrino physics in these sources. We refer the reader to \cite{2018ApJ...852..120B,uribe2019some,universe7010007} for further details. 
%

%%%%%%%%%%%%%%%%%%%%%%%%%%%%%%%%%%%%%%%%%%%%%%%%%%%%%%%%
\subsection{Neutrino Oscillations}
%%%%%%%%%%%%%%%%%%%%%%%%%%%%%%%%%%%%%%%%%%%%%%%%%%%%%%%%

First, we establish the theoretical framework that starts from the setup of the Hamiltonian governing neutrino flavor oscillations. There are four relevant ingredients in such Hamiltonian: geometry, mass content, neutrino content, and neutrino mass hierarchy. The equations governing the system evolution are the quantum Liouville equations
\begin{equation}\label{eq:Liouville}
%\begin{gather}
i\dot{\rho}_{\mathbf{p}} = [H_{\mathbf{p}},\rho_{\mathbf{p}}],\quad 
i\dot{\bar{\rho}}_{\mathbf{p}} = [\bar{H}_{\mathbf{p}},\bar{\rho}_{\mathbf{p}}],
%\end{gather}
\end{equation}
where the Hamiltonian is given by
\begin{widetext}
\begin{subequations}\label{eq:FullHam}
\begin{align}
\mathsf{H}_{\mathbf{p},t}&=\Omega_{\mathbf{p},t}+\sqrt{2}G_{F}\!\!\int\!\!\left( l_{\mathbf{q},t}-\bar{l}_{\mathbf{q},t}\right)\left( 1-\mathbf{v}_{\mathbf{q},t}\cdot\mathbf{v}_{\mathbf{p},t} \right)\frac{d^3\mathbf{q}}{\left(2\pi\right)^3} 
%\nonumber \\
%&\qquad\qquad\qquad\qquad\qquad 
+ \sqrt{2}G_{F}\!\!\int\!\!\left( \rho_{\mathbf{q},t}-\bar{\rho}_{\mathbf{q},t}\right)\left( 1-\mathbf{v}_{\mathbf{q},t}\cdot\mathbf{v}_{\mathbf{p},t} \right)\frac{d^3\mathbf{q}}{\left(2\pi\right)^{3}},\\
\mathsf{\bar{H}}_{\mathbf{p},t}&=-\Omega_{\mathbf{p},t}+\sqrt{2}G_{F}\!\!\int\!\!\left( l_{\mathbf{q},t}-\bar{l}_{\mathbf{q},t}\right)\left( 1-\mathbf{v}_{\mathbf{q},t}\cdot\mathbf{v}_{\mathbf{p},t} \right)\frac{d^3\mathbf{q}}{\left(2\pi\right)^3} 
+ \sqrt{2}G_{F}\!\!\int\!\!\left( \rho_{\mathbf{q},t}-\bar{\rho}_{\mathbf{q},t}\right)\left( 1-\mathbf{v}_{\mathbf{q},t}\cdot\mathbf{v}_{\mathbf{p},t} \right)\frac{d^3\mathbf{q}}{\left(2\pi\right)^{3}}.
\end{align}
\end{subequations}
\end{widetext}
with $\rho_{\mathbf{p}}$ ($\bar{\rho}_{\mathbf{p}}$) the matrix of occupation numbers $\langle a^{\dagger}_{j}a_{i}\rangle_\mathbf{p}(\langle \bar{a}^{\dagger}_{i}\bar{a}_{j}\rangle_\mathbf{p})$ for (anti)neutrinos, for the particle momentum $\mathbf{p}$ and flavors $i,j$. The diagonal elements are the distribution functions $f_{\nu_{i}\left(\bar{\nu}_{i}\right)}\left(\mathbf{p}\right)$. The off-diagonal elements contain information about the flavor \emph{overlapping}. Here, $\Omega_{\mathbf{p}}$ is the matrix of vacuum oscillation frequencies, $l_{\mathbf{p}}$ and $\bar{l}_{\mathbf{p}}$ are occupation number matrices for charged leptons, and $\mathbf{v}_{\mathbf{p}}=\mathbf{p}/ p$ is the normalized particle velocity associated with the particle momentum $\mathbf{p}$. 
\begin{table}
\centering
\caption{Mixing and squared mass differences presented in \cite{ParticleDataGroup:2022pth}. The associated errors are at the 3$\sigma$ level. We recall that $\Delta m^2 = m^{2}_{3}-\left(m^{2}_{2}+m^{2}_{1}\right)/2$, whose sign depends on the hierarchy, i.e., either $m_{1}<m_{2}<m_{3}$ or $m_{3}<m_{1}<m_{2}$.}
\begin{tabular}{ l }
\hline\hline
$ \Delta m^{2}_{21}  = 7.53\,(7.35-7.71)\times 10^{-5}$ eV$^2 $ \\
$\Delta m^{2} (\Delta m^2 > 0) = 2.453\,(2.42-2.486) \times 10^{-3}$ eV$^2$  \\ 
$\vert \Delta m^{2} \vert (\Delta m^2 < 0) = 2.536\,(2.57-2.502) \times 10^{-3}$ eV$^2$  \\ 
$\sin^2\theta_{12} = 0.309\,(0.296-0.322)$ \\
$\sin^2\theta_{23} (\Delta m^2 > 0) = 0.539\,(0.517-0.561)$ \\
$\sin^2\theta_{23} (\Delta m^2 < 0) = 0.546\,(0.525-0.567)$ \\
$\sin^2 \theta_{13} = 0.022\,(0.0213-0.0227)$ \\
\hline
\end{tabular}
\label{tab:massvalues}
\end{table}

Electron neutrinos ($\nu_e$ and $\bar\nu_e$) interact with matter in the accretion zone (e.g., protons, neutrons, electrons, and positrons) via both charged and neutral currents, while $\nu_\mu$, $\nu_\tau$, $\bar\nu_\mu$, and $\bar\nu_\tau$ interact via neutral currents since the accretion zone does not contain muons or tau leptons. The two-flavor approximation is also justified by the strong hierarchy of the squared mass differences $\vert \Delta m^{2}_{13} \vert \approx \vert \Delta m^{2}_{23} \vert \gg \vert \Delta m^{2}_{12} \vert$, and only the mixing angle $\theta_{13}$ is
considered (see Table~\ref{tab:massvalues}). Thus, hereafter, we drop the angle suffix. The states can be divided into electron and non-electron ones, which supports our use of the two-flavor approximation. Therefore, we can write $\rho$ in Eq.~(\ref{eq:Liouville}) with the aid of the Pauli matrices and the polarization vector $\mathsf{P}_\mathbf{p}$ as
\begin{equation}
\small
\rho_{\mathbf{p}}=\left(
 \begin{array}{cc}
  \rho_{ee} & \rho_{ex}\\
  \rho_{xe} & \rho_{xx}\\
   \end{array}\right)_{\mathbf{p}}
   =
 \frac{1}{2}\left(f_{\mathbf{p}}\mathbb{I} +\mathsf{P}_\mathbf{p} \cdot \vec \sigma\right),
	\label{eq:expofrho}
\end{equation}
where $f_{\mathbf{p}}=f_{\nu_e}(\mathbf{p})+f_{\nu_x}(\mathbf{p})$, and the $x$ denotes the non-electron flavor within the present two-flavor approximation. Likewise, one can obtain the corresponding equations for antineutrinos. The polarization vector satisfies
\begin{equation}
\mathsf{P}^{z}_{\mathbf{p}} =f_{\nu_e}(\mathbf{p})-f_{\nu_x}(\mathbf{p}),
\label{eq:pzeta1}
\end{equation}
so it tracks the relative flavor composition. Therefore, by using an adequate normalization of $\rho_{\mathbf{p}}$, it can be used to define the survival and mixing probabilities
\begin{equation}\label{eq:survprobability1}
%\begin{gather}
P_{\nu_{e} \leftrightarrow \nu_{e}} = \frac{1}{2}\left( 1 + \mathsf{P}^{z}_{\mathbf{p}} \right),\quad
P_{\nu_{e} \leftrightarrow \nu_{x}} = \frac{1}{2}\left( 1 - \mathsf{P}^{z}_{\mathbf{p}} \right).
%\end{gather}
\end{equation}

The two-flavor Hamiltonian of Eqs.~(\ref{eq:FullHam}) can be writtent as a sum of three terms
\begin{equation}
\mathsf{H} = \mathsf{H}_{\mbox{\footnotesize{vac}}} + \mathsf{H}_{\mbox{\footnotesize{m}}} + \mathsf{H}_{\nu\nu}.
\label{neutrinohamiltonian}
\end{equation}

The first term is the vacuum Hamiltonian~\cite{Qian:1994wh}
\begin{equation}
\mathsf{H}_{\mbox{\footnotesize{vac}}} =\frac{\omega_\mathbf{p}}{2}
\left(
 \begin{array}{cc}
  -\cos 2\theta & \sin 2\theta\\
  \sin 2\theta & \cos 2\theta \\
   \end{array}\right)
   =\frac{\omega_\mathbf{p}}{2} \mathbf{B}\cdot \vec{\sigma},
	\label{Hvacuum}
\end{equation}
where $\mathbf{B}=(\sin2\theta,0,-\cos 2 \theta)$ and $\omega_\mathbf{p} = \Delta m^2/2p$. 

The matter Hamiltonian is
\begin{equation}
\mathsf{H}_{\mbox{\footnotesize{m}}} =
\frac{\lambda}{2}\left(
 \begin{array}{cc}
  1 & 0\\
  0 & -1 \\
   \end{array}\right)
   =\frac{\lambda}{2} \mathbf{L} \cdot \vec{\sigma},
	\label{Hmatter}
\end{equation}
with $\lambda = \sqrt{2}G_{F}\left(n_{e^-} - n_{e^+}\right)$ as the matter potential and $\mathbf{L}=(0,0,1)$. We have assumed that electrons form an isotropic gas, so the vector $\mathbf{v}_{\mathbf{q}}$ is distributed uniformly on the unit sphere, and the average of $\mathbf{v}_\mathbf{q}\cdot\mathbf{v}_\mathbf{p}$ averages vanishes.

The above simplification is not possible with the final term since the net neutrino and antineutrino fluxes are non-zero, so $\mathbf{v}_\mathbf{q}\cdot\mathbf{v}_\mathbf{p}$ does not vanish. We can use Eq.~(\ref{eq:expofrho}) and obtain \cite{1992PhLB..287..128P,2016arXiv160704671Z,2016PhRvD..93d5021M}
\begin{equation}
\mathsf{H}_{\nu\nu} = \sqrt{2}G_{F}\left[ \int\!\! \left(1- \mathbf{v}_{\mathbf{q}}\cdot\mathbf{v}_{\mathbf{p}}\right) \left(\mathsf{P}_\mathbf{q}-\bar{\mathsf{P}}_\mathbf{q}\right)\frac{d^3\mathbf{q}}{\left(2\pi\right)^3}\right]\cdot \vec{\sigma}.
\label{Hnunu}
\end{equation}

Introducing each term in Eqs.~(\ref{eq:FullHam}), and using the commutation relations of the Pauli matrices, we find from Eqs.~(\ref{eq:Liouville}) the equations of neutrino (and antineutrino) oscillations for each momentum mode $\mathbf{p}$:
\begin{subequations}
\begin{align}
\begin{split}
\dot{\mathsf{P}}_\mathbf{p} = \Biggl[ \omega_\mathbf{p} \mathbf{B} + &\lambda \mathbf{L} +  \sqrt{2}G_{F}  \int\!\! \left(1- \mathbf{v}_{\mathbf{q}}\cdot\mathbf{v}_{\mathbf{p}}\right) \\
&\left(\mathsf{P}_\mathbf{q}-\bar{\mathsf{P}}_\mathbf{q}\right)\frac{d^3\mathbf{q}}{\left(2\pi\right)^3} \Biggr] \times \mathsf{P}_\mathbf{p},
\end{split}\\
\begin{split}
\dot{\bar{\mathsf{P}}}_\mathbf{p} = \Biggl[ -\omega_\mathbf{p} \mathbf{B} &+ \lambda \mathbf{L} +  \sqrt{2}G_{F}\!\!  \int\!\! \left(1- \mathbf{v}_{\mathbf{q}}\cdot\mathbf{v}_{\mathbf{p}}\right)\\
&\left(\mathsf{P}_\mathbf{q}-\bar{\mathsf{P}}_\mathbf{q}\right)\frac{d^3\mathbf{q}}{\left(2\pi\right)^3}\Biggr]\times \bar{\mathsf{P}}_\mathbf{p}.
\end{split}
\end{align}
\label{eq:Hnu1}
\end{subequations}
%

%%%%%%%%%%%%%%%%%%%%%%%%%%%%%%%%%%%%%%%%%%%%%%%%%%%%
\subsection{Neutrino Emission in the Hypercritical Accretion onto the NS}\label{subsec:spherical}
%%%%%%%%%%%%%%%%%%%%%%%%%%%%%%%%%%%%%%%%%%%%%%%%%%%%

In the accretion process, the infalling material compresses, so it becomes sufficiently hot to produce thermally $e^+e^-$ pairs whose annihilation leads to a high neutrino flux. Neutrinos take away most of the infalling matter's gravitational energy gain, reducing its entropy and allowing it to be incorporated into the NS. Near the NS surface, the matter temperature $T$ is so high that it is in a non-degenerate, relativistic, hot plasma state. Under these conditions, the most efficient neutrino emission channel is the $e^+e^-$ pair annihilation process \cite{2016ApJ...833..107B}. The neutrino emissivity can be approximated by
% 
%\begin{widetext}
\begin{align}\label{eq:approximationyakovlev}
\varepsilon^{(m)}_{i} &\approx \frac{2G^{2}_{F}\left(T\right)^{8+m}}{9\pi^{5}}C^{2}_{+,i}\left[\mathcal{F}_{m+1}\left(\eta_{e^{+}}\right)\mathcal{F}_{1}\left(\eta_{e^{-}}\right) \right.\nonumber \\
&+ \left.\mathcal{F}_{m+1}\left(\eta_{e^{-}}\right)\mathcal{F}_{1}\left(\eta_{e^{+}}\right)\right],
\end{align}
%\end{widetext}
%
where $G_F$ is the weak interaction Fermi constant,  $C^{2}_{+,i} = C^{2}_{V_i} + C^{2}_{A_i}$, with $C_{V_e} = 2\sin^{2}\theta_W + 1/2$, $C_{A_e}=1/2$, $C_{V_\mu}=C_{V_\tau}=C_{V_e}-1$, $C_{A_\mu}=C_{A_\tau}=C_{A_e}-1$, $\sin^{2}\theta_W\approx 0.231$, $\mathcal{F}_{m}\left(\eta \right) = \int_{0}^{\infty}dx x^{m}/[1+\exp{(x-\eta)}]$ is the generalized Fermi function (see, e.g., \cite{2018ApJ...852..120B}), with $\eta$ the degeneracy parameter and $m$ the index of the Fermi functions. For $m=0$ and $m=1$, Eq. (\ref{eq:approximationyakovlev}) gives, respectively, the neutrino and antineutrino (of flavor $i$) number emissivity and energy emissivity. From Eq.~(\ref{eq:approximationyakovlev}), we find
\begin{subequations}
\begin{gather}
\frac{\varepsilon^{(0)}_{i}}{2} = n^{C}_{\nu_{i}}=n^{C}_{\bar{{\nu_{i}}}}, \;\, F^{C}_{\nu_{i}} = F^{C}_{\bar{\nu_{i}}} \;\,\, \forall i\, \in \{ e,\mu,\tau \},\\
\frac{\varepsilon^{(0)}_{e}}{\varepsilon^{(0)}_{x}} = \frac{n^{C}_{\nu_e}}{n^{C}_{\nu_x}} =\frac{n^{C}_{\bar{\nu}_e}}{n^{C}_{\bar{\nu}_x}} = \frac{F^{C}_{\nu_e}}{F^{C}_{\nu_x}} = \frac{F^{C}_{\bar{\nu}_e}}{F^{C}_{\bar{\nu}_x}}  \approx \frac{7}{3},
\end{gather}\label{eq:neutrinoratio}\end{subequations}
where the density and flux at creation are, respectively, $n^{c}_{\nu_{i}(\bar{{\nu_{i}}})}$ and $F^{c}_{\nu_{i}(\bar{\nu_{i}})}$. Eqs~(\ref{eq:neutrinoratio}) show that in the present case, of the total number of neutrinos+antineutrinos, $70\%$ are electron, $30\%$ are non-electron, while the number of neutrinos (of all flavors) equals the number of antineutrinos.

Adding all flavors in the case $m=1$ and $\eta_{e^{\pm}}=0$, Eq.~(\ref{eq:approximationyakovlev}) reduces to
\begin{equation}\label{eq:L_neutrinos}
\epsilon_{e^{-}\!e^{+}} \approx 8.69\times 10^{30}\left(\frac{T}{1\,{\rm MeV}}\right)^9\,\, {\rm MeV}\,{\rm cm}^{-3}\,{\rm s}^{-1},
\end{equation}
where $\epsilon_{e^{-}\!e^{+}}=\varepsilon^{(1)}_{e} + \varepsilon^{(1)}_{x}$ is the total emissivity. The average neutrino or antineutrino energy of flavor $i$ can be estimated as $\langle E_{\nu_{i}} \rangle = \varepsilon^{(1)}_{i}/\varepsilon^{(0)}_{i}$. In particular, we find for each flavor
\begin{equation}
\langle E_{\nu}  \rangle = \langle E_{\bar{\nu}}  \rangle \approx 4.1\,T.
\end{equation}\label{eq:neutrinotwomoments}

Using Eq.~(\ref{eq:L_neutrinos}), we define an effective neutrino emission region \cite{2018ApJ...852..120B}
\begin{equation}
\Delta r_{\nu} = \frac{\epsilon_{e^{-}\!e^{+}}}{\nabla \epsilon_{e^{-}\!e^{+}}} = \approx 0.08R_{\rm NS},
\label{neutrinoshell}
\end{equation}
where $R_{\rm NS}$ is the NS radius. The above implies that the neutrino emission region is thin, so we consider it a spherical shell and apply the single-angle approximation~\cite{Duan:2006an,Dasgupta:2007ws}\footnote{The \emph{multi-angle} terms lead to kinematic decoherence~\cite{Hannestad:2006nj,Raffelt:2007yz,2007JCAP...12..010F}.}. The above simplifies the last term in Eq.~(\ref{eq:Hnu1}), so the potentials simplify to the expressions
\begin{subequations}
\begin{align}
\begin{split}
\omega_{p,r} &= \frac{\Delta m^{2}}{2p\langle v_{r} \rangle},
\label{eq:vacuumpotential}    
\end{split}\\
\begin{split}
\lambda_{r} &= \sqrt{2}G_{F}\left(n_{e^{-}}-n_{e^{+}}\right)\frac{1}{\langle v_{r} \rangle},
\label{eq:matterpotential}
\end{split}\\
\begin{split}
\mu_{r} &=\frac{ \sqrt{2}G_{F}}{2}\left(\sum_{i\in\{e,x\}}\!n^{C}_{\nu_{i}\bar{\nu}_{i}}\right)\left( \frac{R_{\rm NS}}{r} \right)^{2}\left( \frac{1 - \langle v_{r} \rangle^{2}}{\langle v_{r} \rangle} \right),
\label{eq:neutrinopotential}    
\end{split}
\end{align}
\end{subequations}
where
\begin{equation}
\langle v_{r} \rangle = \frac{1}{2}\left[ 1+\sqrt{1 - \left(\frac{R_{\rm NS}}{r}\right)^{2}} \right],
\label{eq:averageradialvelocity}
\end{equation}
being $r$ the radial distance from the NS center. Here, $\omega_{p,r}$ is the vacuum potential, $\lambda_{r}$ is the matter potential, and $\mu_{r}$ is the self-interaction potential.

\begin{figure*}
\centering
\includegraphics[width=0.5\hsize,clip]{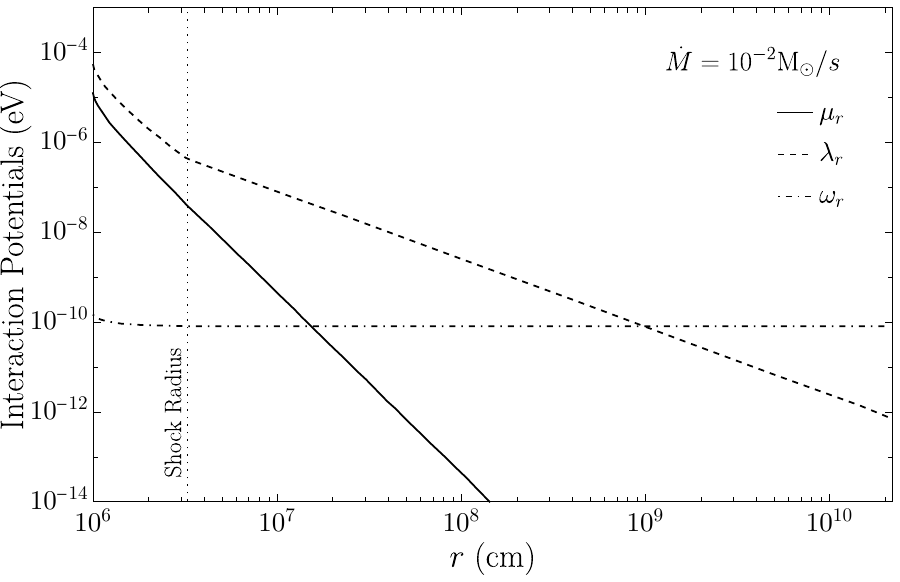}\includegraphics[width=0.5\hsize,clip]{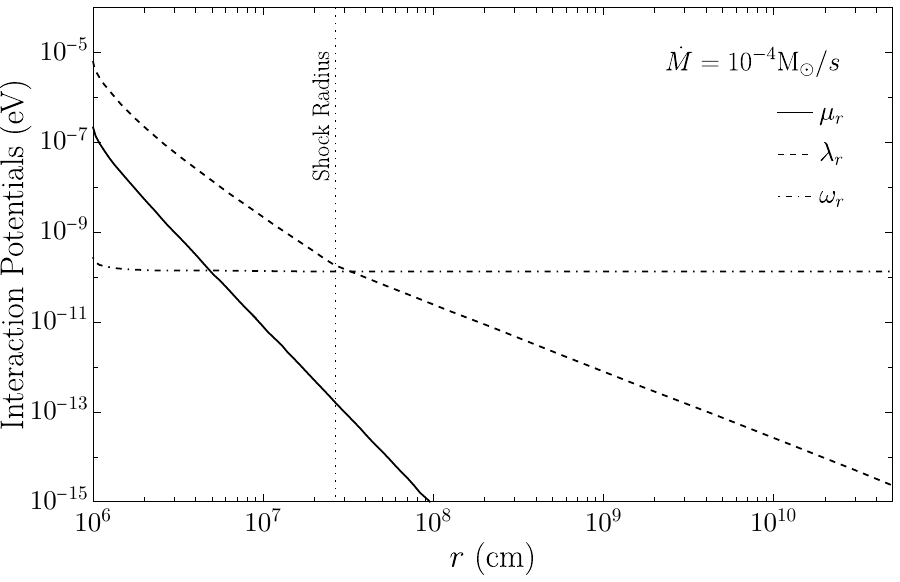}
\caption{Oscillation potentials from the NS surface to the accretion (Bondi-Hoyle) radius, for the values of $\dot{M}$ in Table \ref{tab:tab1}. Figure reproduced from~\cite{2018ApJ...852..120B}.}
\label{fig:potentials}
\end{figure*}
Table~\ref{tab:tab1} lists the thermodynamic properties of the accreting matter at the NS surface, obtained from Eq.~(\ref{eq:approximationyakovlev}) and the hydrodynamic simulations in~\cite{2016ApJ...833..107B}.
\begin{table*}
\caption{Accretion parameters at the NS surface. $n^{C}$ and $F^{C}$ represent the neutrino number density and flux. Reproduced from~\cite{2018ApJ...852..120B}.}
\begin{tabular}{c c c c c c c c c c c c c c c}
\hline
  $\dot{M}$ & $\rho$ & $T$ & $\eta_{e^{\mp}}$ & $n_{e^{-}}\!-n_{e^{+}}$ & $T_{\nu\bar{\nu}}$ & $\langle E_\nu \rangle$ & $F^{C}_{\nu_e,\bar{\nu}_e}$ & $F^{C}_{\nu_x,\bar{\nu}_x}$ & $n^{C}_{\nu_{e}\bar{\nu}_{e}}$ & $n^{C}_{\nu_{x}\bar{\nu}_{x}}$\ & $\sum_{i}\,n^{C}_{\nu_{i}\bar{\nu}_{i}}$ \\ 
	$(M_\odot$~s$^{-1}$) & (g~cm$^{-3})$ & (MeV) &  & (cm$^{-3}$) & (MeV) & (MeV) & (cm$^{-2}$s$^{-1}$) & (cm$^{-2}$s$^{-1}$) & (cm$^{-3})$ & (cm$^{-3})$ & (cm$^{-3})$ \\ \hline
  $10^{-2}$ & $7.54\times10^{8}$ & 7.13 & $\mp 0.057$ & $2.27\times10^{32}$ & 8.08 & 29.22 & $7.92\times 10^{41}$  & $3.39\times 10^{41}$ & $5.28\times10^{31}$ & $2.26\times10^{31}$ & $7.54\times10^{31}$ \\
  $10^{-3}$ & $2.48\times10^{8}$ & 5.54 & $\mp 0.082$ & $7.65\times10^{31}$ & 6.28 & 22.70 & $1.04\times 10^{41}$ & $4.51\times 10^{40}$ & $7.00\times10^{30}$ & $3.00\times10^{30}$ & $1.00\times10^{31}$ \\
  $10^{-4}$ & $8.66\times10^{7}$ & 4.30 & $\mp 0.111$ & $2.61\times10^{31}$ & 4.87 & 17.62 &  $1.39\times 10^{40}$ & $5.94\times 10^{39}$ & $9.24\times10^{29}$ & $3.96\times10^{29}$ & $1.32\times10^{30}$ \\
  $10^{-5}$ & $3.10\times10^{7}$ & 3.34 & $\mp 0.147$ & $9.56\times10^{30}$ & 3.78 & 13.69 & $1.84\times 10^{39}$ & $7.87\times 10^{38}$ & $1.23\times10^{29}$ & $5.20\times10^{28}$ & $1.75\times10^{29}$ \\ \hline  
\end{tabular}
\label{tab:tab1}
\end{table*}

Figure~\ref{fig:potentials} shows the effective potentials for $\dot{M}= 10^{-2} M_\odot$~s$^{-1}$ and $10^{-4} M_\odot$~s$^{-1}$. We have used the neutrino energy given by the average in Table \ref{tab:tab1}. Different values of $\dot{M}$ can be interpreted either as the evolution of a time-varying accretion rate or as peak accretion rates occurring in CO-NS binaries of different orbital periods. The value of $\dot{M}$ fixes the temperature and density at the NS surface, the effective potentials, and the initial neutrino and antineutrino flavor ratios. We limit ourselves to the conditions reported in Table 1 of \cite{2016ApJ...833..107B}, i.e., accretion rates in the range $\sim 10^{-2}$--$10^{-4} M_\odot$~s$^{-1}$.

\begin{figure*}
\includegraphics[width=0.5\hsize,clip]{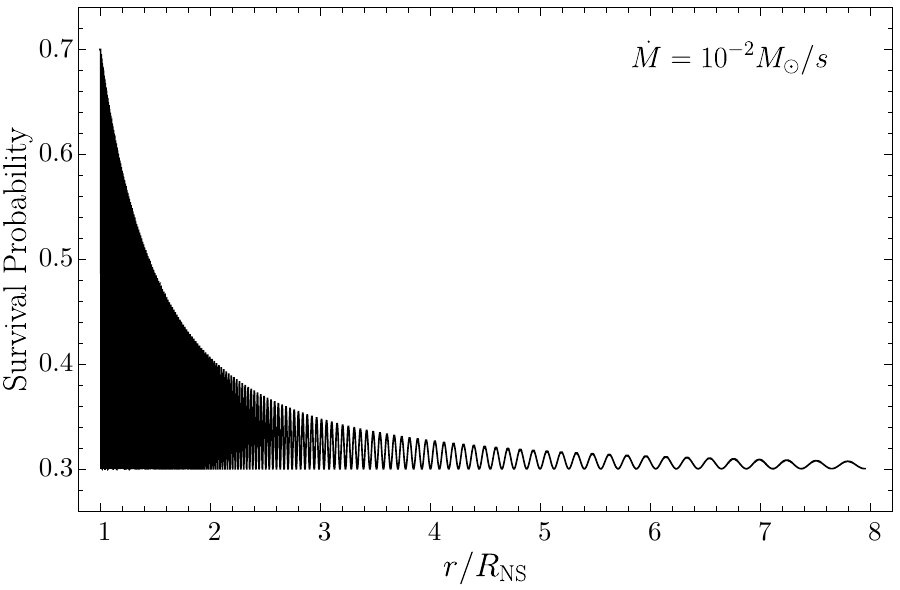}\includegraphics[width=0.5\hsize,clip]{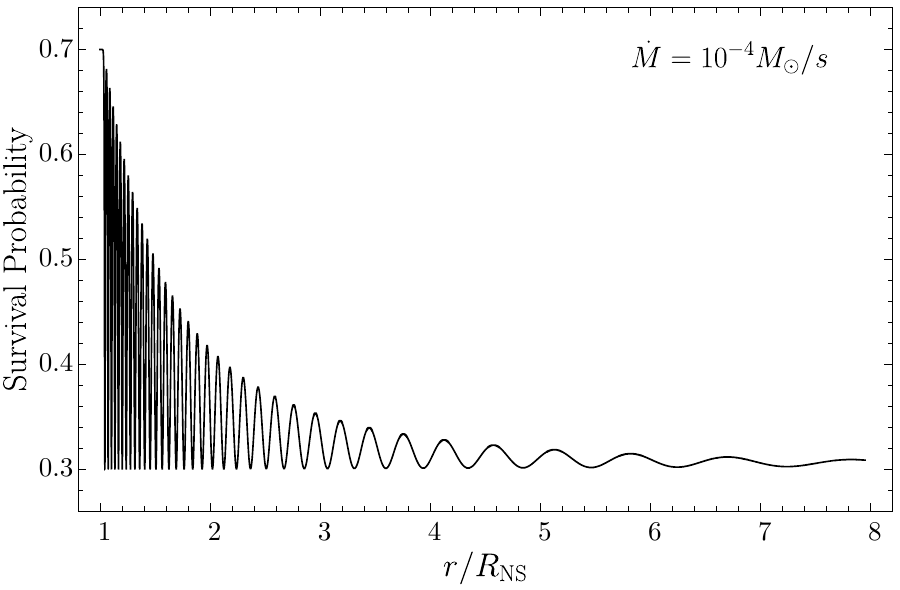}\\
\includegraphics[width=0.5\hsize,clip]{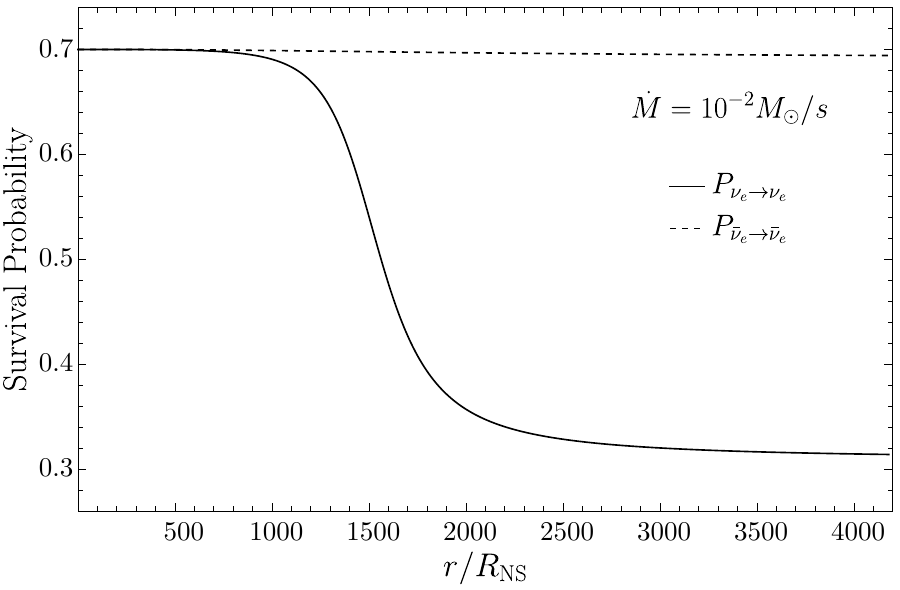}\includegraphics[width=0.5\hsize,clip]{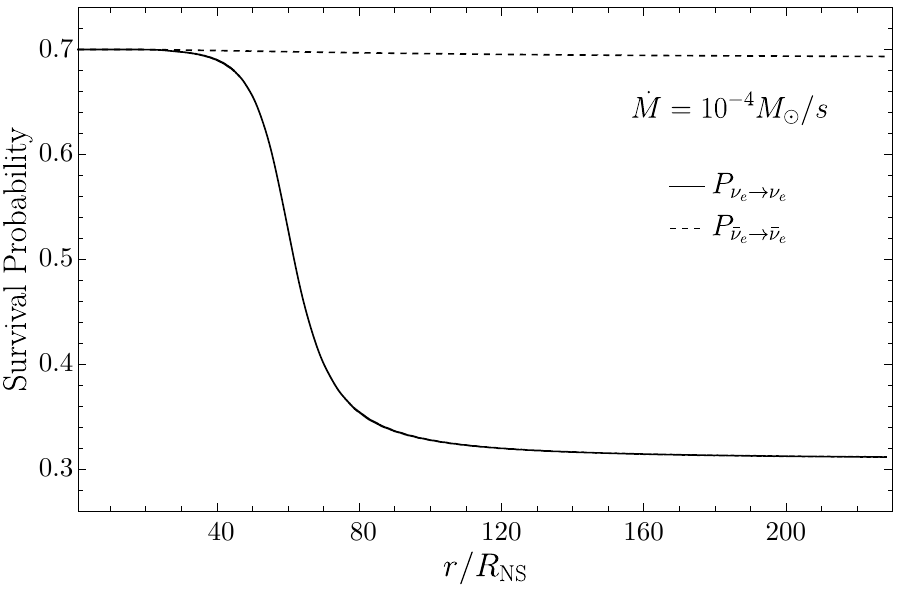}
\caption{Flavor content for inverted (top panels) and normal hierarchy (bottom panels) for selected values of $\dot{M}$. In the inverted hierarchy case, the curves for electron neutrino and antineutrino coincide. Reproduced from~\cite{2018ApJ...852..120B}.} 
\label{fig:singleangle}
\end{figure*}

In Fig.~\ref{fig:singleangle}, for $\dot{M}=10^{-2}$ and $10^{-4}$ $M_{\odot}$~s$^{-1}$, we show the solution of Eqs.~(\ref{eq:Hnu1}) for both mass hierarchies and a monochromatic neutrino spectrum given by the average neutrino energy. For the inverted hierarchy, the neutrino and antineutrino survival probabilities are equal because the matter and self-interaction potentials are much larger than the vacuum potential. The antineutrino flavor ratios remain unchanged in the normal hierarchy, but the electron neutrino flavor ratios change when $\lambda_{r} \sim \omega_{r}$. From these results, we estimate an oscillation length
\begin{equation}
t_{\rm osc} \approx (50-1000) \,\,{\rm m},
\label{length}
\end{equation}
which agree with previous estimates in \cite{Hannestad:2006nj,Raffelt:2007yz}.  \citet{Hannestad:2006nj,Raffelt:2007yz,2007JCAP...12..010F} have argued that multi-angle effects lead to kinematic decoherence in both mass hierarchies, while \citet{EstebanPretel:2007ec} discussed decoherence due to neutrino flavor asymmetry. They concluded that when the difference between the number of neutrinos and antineutrinos is $\gtrsim 25\%$ than the total number of neutrinos, decoherence becomes irrelevant.
\begin{table*}
\caption{Comparison between initial and final (anti)neutrino flavor content. The total neutrino number density is $n=2\sum_{i}n_{\nu_{i}}$. Figure reproduced from~\cite{2018ApJ...852..120B}.}
\begin{tabular}{c c c c c c c c c c c c}
\hline
  & $n^{0}_{\nu_{e}}/n$ & $n^{0}_{\bar{\nu}_{e}}/n$ & $n^{0}_{\nu_{x}}/n$ & $n^{0}_{\bar{\nu}_{x}}/n$ & $n_{\nu_{e}}/n$ & $n_{\bar{\nu}_{e}}/n$ & $n_{\nu_{x}}/n$ & $n_{\bar{\nu}_{x}}/n$ \\ 
 \hline\hline
    Normal Hierarchy & $\frac{1}{6}$ & $\frac{1}{6}$ & $\frac{1}{3}$ & $\frac{1}{3}$ & $\frac{1}{3}$ & $\frac{1}{6} + \frac{1}{6}\sin^{2}\theta_{12}$ & $\frac{1}{6}$ & $\frac{1}{3}-\frac{1}{6}\sin^{2}\theta_{12}$ \\ \hline
    Inverted Hierarchy & $\frac{1}{6}$ & $\frac{1}{6}$ & $\frac{1}{3}$ & $\frac{1}{3}$ & $\frac{1}{6} + \frac{1}{6}\cos^{2}\theta_{12}$ & $\frac{1}{3}$ & $\frac{1}{3}-\frac{1}{6}\cos^{2}\theta_{12}$ & $\frac{1}{6}$ \\ \hline    
\end{tabular}
\label{tab:tabfluxes}
\end{table*}

Since neutrinos and antineutrinos are created in equal amounts, bipolar oscillations are inevitable. The oscillation length of the bipolar behavior follows Eq.~(\ref{length}). However, the oscillation length is a function of the neutrino energy, and averaging over the neutrino energy spectrum should lead to flavor equipartition within a few oscillation cycles~\cite{Raffelt:2007yz}. Consequently, we extend the two-flavor approximation to three-flavors, i.e.,
\begin{equation}
n_{\nu_e(\bar{\nu}_e)}=n_{\nu_\mu(\bar{\nu}_\mu)}=n_{\nu_\tau(\bar{\nu}_\tau)}.
\label{eq:proportion}
\end{equation}

Figure~\ref{fig:potentials} shows that at distances $r \gtrsim 1.08R_{\rm NS}$, the self-interaction potential $\mu_r$ in Eq.~(\ref{eq:neutrinopotential}) decays faster than the matter potential $\lambda_r$ in Eq.~(\ref{eq:matterpotential}), so the latter becomes responsible for the oscillations. The matter potential inhibits neutrino oscillations, freezing the neutrino content. Yet, we expect the neutrino content to change due to the Mikheyev-Smirnov-Wolfenstein (MSW) effect~\cite{1978PhRvD..17.2369W, Mikheyev1986}. That is, whenever the matter potential satisfies the resonance condition
\begin{equation}
\lambda_r \sim \omega_r.
\label{resonances}
\end{equation}

Let $F^{0}$ and $F$ represent the neutrino flux after bipolar oscillations and after the MSW effect, respectively. Then, we have 
\begin{subequations}
\begin{gather}
F_{\nu_e}=P_{\nu_e\to \nu_e}F^0_{\nu_e}+\left[1-P_{\nu_e\to \nu_e}\right]F^0_{\nu_x},\\
F_{\bar{\nu}_e}=P_{\bar{\nu}_e\to \bar{\nu}_e}F^0_{\bar\nu_e}
+\left[1-P_{\bar\nu_e\to \bar\nu_e}\right]F^0_{\bar\nu_x},
\end{gather}\label{eq:fluxatearth}
\end{subequations}
where $P$ is the survival probability for the transition through the MSW region.

To calculate the fluxes beyond the MSW region, we use the results in~\cite{Fogli:2003dw,2018ApJ...852..120B}. For normal hierarchy,
\begin{equation}
P_{\nu_e\to \nu_e}= X \sin^2\theta_{12},\quad P_{\bar\nu_e\to \bar\nu_e}= \cos^2\theta_{12},
\end{equation}
and, for inverted hierarchy,
\begin{equation}
P_{\nu_e\to \nu_e}=\sin^2\theta_{12},\quad
P_{\bar\nu_e\to \bar\nu_e}= X \cos^2\theta_{12},
\end{equation}
where $X$ is given by \cite{petcov1987non,Fogli:2003dw,Kneller:2005hf}
\begin{equation}
X = \frac{\exp(2\hat{r}\hat{k} \cos 2 \theta_{13}) - 1}{\exp(2 \hat{r}\hat{k}) - 1},
 \label{eqX}
 \end{equation}
being $\hat{r}=r$ such that Eq.~(\ref{resonances}) is satisfied and  
\begin{equation}
\frac{1}{\hat{k}} = \left\vert \frac{d \ln \lambda _{r}}{dr}\right\vert_{r=\hat{r}}.
\label{eqk}
\end{equation}
The function $X$ measures the speed of changes of the matter potential. $X\to 0$ and $X\to 1$ represent adiabatic and non-adiabatic changes, respectively. The MSW resonance~\cite{1978PhRvD..17.2369W, Mikheyev1986} happens far from the emission region, so $X \to 0$ (see Fig.~\ref{fig:potentials}). We can obtain the final neutrino fluxes by applying this condition to Eq.~(\ref{eq:fluxatearth}). Table~\ref{tab:tabfluxes} shows a comparison between the neutrino content at the NS surface and after the MSW effect.

In the overall picture, the conditions of the physical system imply that neutrinos radiated from the NS surface experience bipolar oscillations first, leading to flavor equipartition. After the neutrino density decreases due to flux dilution, the MSW resonance dictates the flavor behavior. Consequently, the flavor content exiting the system differs from the flavor content at the NS surface. In particular, the initial fraction in Eq.~(\ref{eq:neutrinoratio}) becomes
\begin{equation}
\frac{n_{\nu_e}}{n_{\nu_x}} = \frac{11}{9},\qquad \frac{n_{\nu_e}}{n_{\nu_x}} = \frac{31}{19},
\label{eq:fratio_normal}
\end{equation}
for normal and inverted hierarchy, respectively. We refer to \cite{2018ApJ...852..120B} for further details.
%

%%%%%%%%%%%%%%%%%%%%%%%%%%%%%%%%%%%%%%%%%%%%%%%%%%%%
\subsection{Neutrino Emission from the Accretion Disk Around the Newborn BH}\label{subsec:disks}
%%%%%%%%%%%%%%%%%%%%%%%%%%%%%%%%%%%%%%%%%%%%%%%%%%%%

% 
\begin{figure*}
\centering
\includegraphics[width=0.49\hsize,clip]{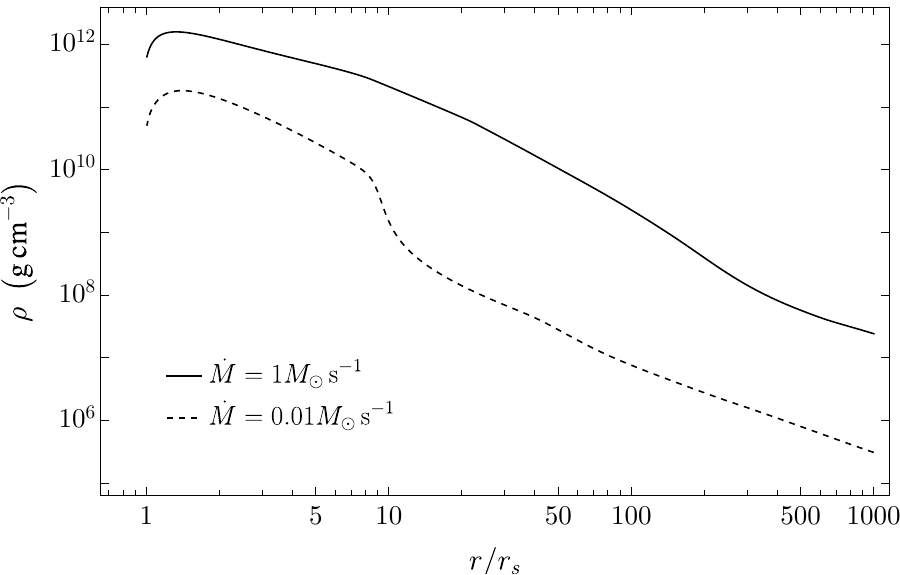}\includegraphics[width=0.49\hsize,clip]{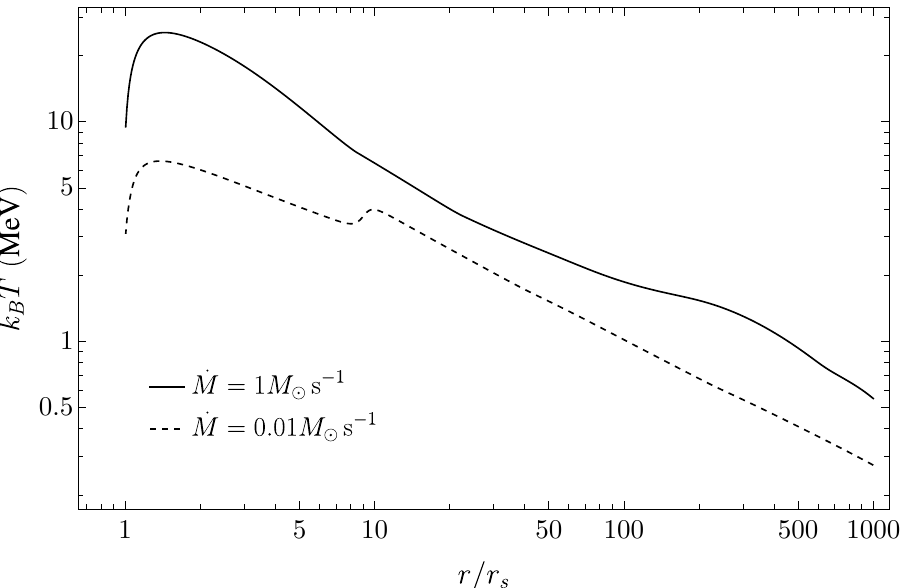}\\
\includegraphics[width=0.49\hsize,clip]{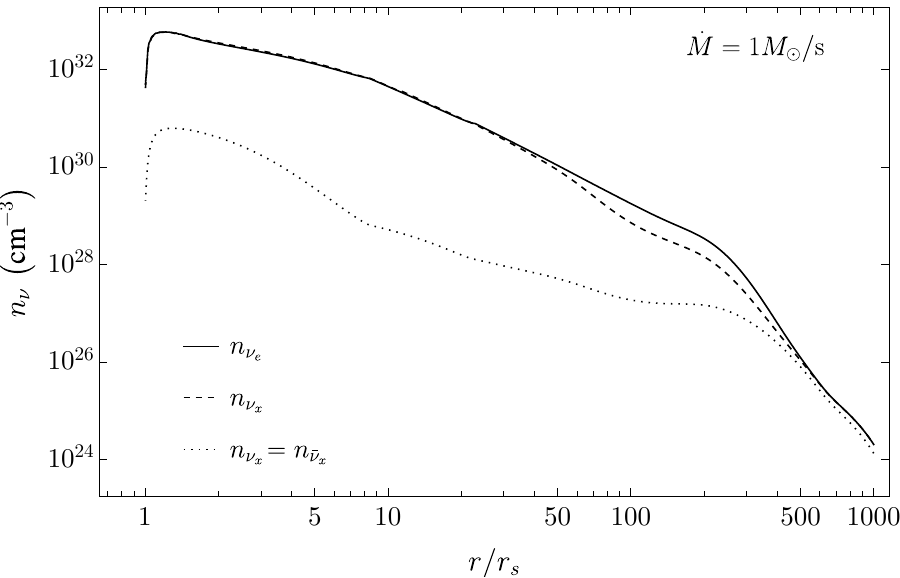}\includegraphics[width=0.49\hsize,clip]{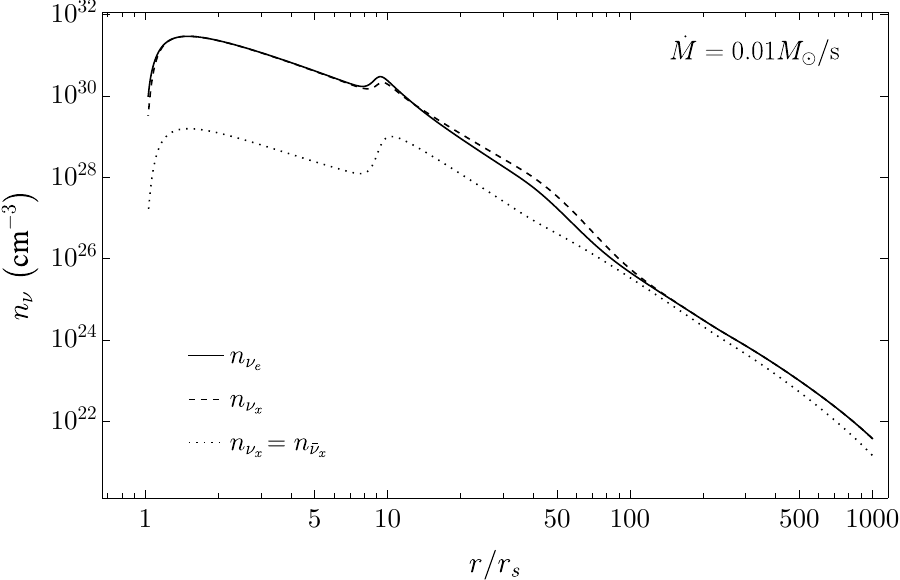}
\includegraphics[width=0.49\hsize,clip]{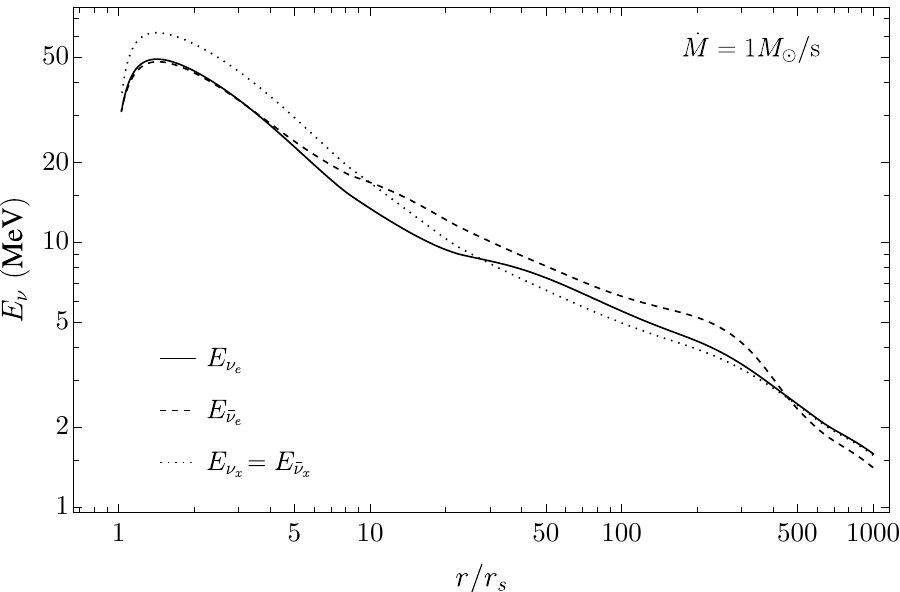}\includegraphics[width=0.49\hsize,clip]{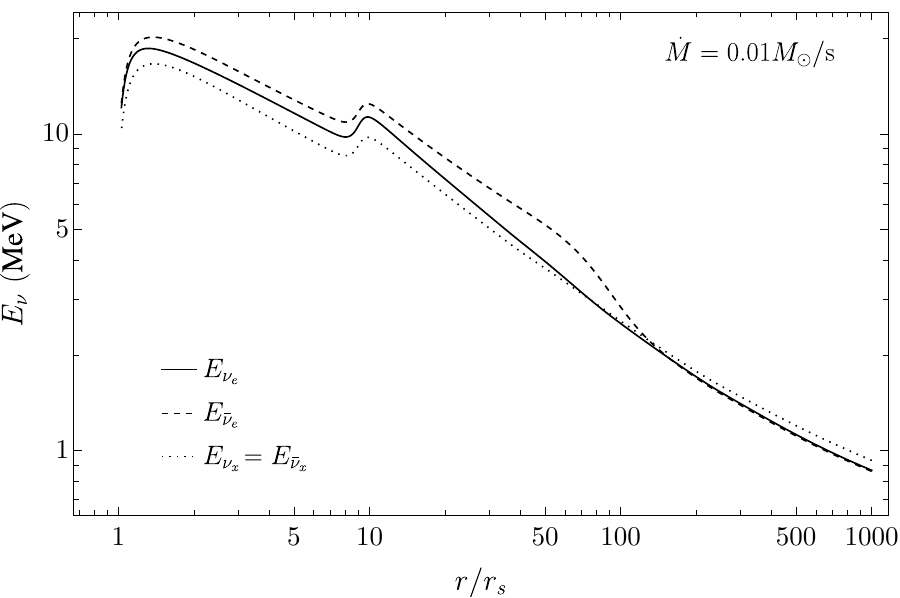}\caption{Density and temperature of the accretion disk (top panels), neutrino number density (middle panels), and neutrino energy density (bottom panels) for selected values of the accretion rate and as functions of the coordinate $r$ normalized by the Schwarzschild radius $r_s$. The disk parameters are $M=3M_{\odot}$, $a = 0.95$ and $\alpha = 0.01$. Reproduced from~\cite{universe7010007}.} 
\label{fig:Disks}
\end{figure*}
In a later stage of the BdHN scenario, the ejecta may circularize and form an accretion disk around the new Kerr BH~\cite{2019ApJ...871...14B}. Our purpose now is to introduce the dynamics of neutrino oscillations in such accretion disks. We adopt the simple yet robust neutrino-cooled accretion disk (NCAD) model~\cite{universe7010007, 1973A&A....24..337S, 1973blho.conf..343N, 1974ApJ...191..499P, krolik1999active,abramowicz1998theory, 2007ApJ...657..383C, LIU20171}. NCADs can reach densities as high as $\sim 10^{10}$---$10^{13}$~g~cm$^{-3}$ and temperatures as high as $\sim 10^{10}$--$10^{11}$~K around the inner disk radius. In such an environment, (anti)neutrinos are created in large amounts by $e^{-}e^{+}$ annihilation, nucleon-nucleon bremsstrahlung, and URCA processes. Once (anti)neutrinos escape the disk, they annihilate around the Kerr BH, creating an $e^{-}e^{+}$ plasma, making them of interest for GRB physics. The system involves neutrinos propagating through dense media, which requires an analysis of neutrino oscillations. The energy density of the electron-positron plasma depends on the (anti)neutrino energy density and flavor content inside and above the accretion disk.

To include neutrino oscillations in the accretion disk model, we need to determine the initial flavor content and the oscillation potentials necessary to solve Eq.~(\ref{eq:Hnu1}). This requires solving the hydrodynamic model without flavor oscillations as a first approximation. The most common NCAD model is the vertically integrated, thin, nearly geodesic accretion disk around a Kerr BH~\cite{1973blho.conf..343N, 1974ApJ...191..499P, 1974ApJ...191..507T}. The disk lies in and around the equatorial plane ($\theta \sim \pi/2$), where the spacetime is represented by the Kerr metric $g_{\mu\nu}$ in cylindrical Boyer-Lindquist coordinates $x^\mu = (t, r, z, \phi)$. Here, $z$ measures the vertical distance to the equatorial plane. The components of the fluid four-velocity are $u^\mu = u^t(1, u^r/u^t, 0, \Omega)$, where $\Omega=u^{\phi}/u^{t}$ is the angular velocity of circular geodesics. The hydrodynamic equations involve the fluid density $\rho$, energy density $U$, and pressure $P$, as measured by a comoving observer. The equations represent the conservation of energy, lepton number, mass, and vertical hydrostatic equilibrium. These are, respectively,
\begin{equation}
u^{r}\left[\partial_{r}\left(HU\right)-\frac{U + P}{\rho}\partial_{r}\left(H\rho\right)\right] =2\eta H\sigma^{r\phi}\sigma_{r\phi} - H\epsilon,
\label{eq:energycon}
\end{equation}
\begin{equation}
u^{r}H\left[n_{B}\partial_{r}Y_{e} + \partial_{r} \!\! \sum_{\ell \in \{e,x\}} \! \left(n_{\nu_{\ell}}\! - n_{\bar{\nu}_{\ell}}\right)\right] = \! \sum_{\ell \in \{e,x\}} \! \left( \dot{n}_{\bar{\nu}_{\ell}}\! - \dot{n}_{\nu_{\ell}} \right),
\label{eq:leptonconservation}
\end{equation}
\begin{equation}
4\pi Hr\rho u^{r} = - \dot{M},
\label{eq:masscon}
\end{equation}
\begin{equation}
P=\frac{1}{3}\rho H^{2}\! \left. {R^{z}}_{tzt}\, \right|_{_{z=0}},
    \label{eq:press}
\end{equation}
where $\boldsymbol{\sigma}$ and $\boldsymbol{R}$ are the shear and Riemann tensors, $H$ is half the disk's thickness, $H\epsilon$ is the average energy flux leaving the disk's surface, $n_{B}$ is the baryon number density, $Y_e$ is the electron fraction, $n_{\nu_{j}\left(\bar{\nu}_{j}\right)}$ is the (anti)neutrino number density, and $\dot{n}_{\nu_{j}\left(\bar{\nu}_{j}\right)}$ is the (anti)neutrino number flux. The first and second terms on the r.h.s. of Eq.~(\ref{eq:energycon}) are the heating and cooling rates, $F_{\text{heat}}$ and $F_{\text{cool}}$, and  $\eta =  2\alpha H (P\rho)^{1/2}$, where $\alpha$ is the (turbulent) viscosity factor~\cite{1973A&A....24..337S}. At some distance, $r=r_{\rm ign}$, called ignition radius, the cooling and heating rates become comparable ($2F_{\nu} \sim F_{\rm heat}$). Inside this radius, the accretion disk is sensible to neutrino physics.

The main contribution to the disk's mass-energy comes from protons, neutrons, and ions, so $\rho = \rho_{\textrm{B}} = n_{B}m_{B}$, where $m_{B}$ is the baryon mass, and $n_{B}Y_{e} = n_{e^{-}} - n_{e^{+}}$ by local charge neutrality. The baryonic mass follows the Maxwell-Boltzmann distribution, and nuclear statistical equilibrium determines its composition. Other particles in the disk (neutrinos, electrons, photons, and their respective antiparticles) follow their usual equations of the state except for neutrinos. When the disk's temperature and density are high enough to trap neutrinos and keep them in thermal equilibrium, their pressure, energy density, and number density ($n_{\nu},U_{\nu},P_{\nu}$) follow the Fermi-Dirac distribution. If neutrinos are not trapped, the Fermi-Dirac distribution does not apply. However, at any point in the disk, we can estimate the (anti)neutrino number and energy density from the emission rates of the processes that create them and model the transition between both regimes with the formula~\cite{2007ApJ...657..383C}
\begin{equation}
\langle E_{\nu} \rangle = \left(1 - w_{\nu}\right)\frac{U^{\textrm{free}}_{\nu}}{n^{\textrm{free}}_{\nu}} + w_{\nu}\frac{U^{\textrm{trapped}}_{\nu}}{n^{\textrm{trapped}}_{\nu}},
    \label{eq:neutrinoaverageenergy}
\end{equation}
where
\begin{equation}
w_{\nu} = \frac{U^{\textrm{free}}_{\nu}}{U^{\textrm{free}}_{\nu} + U^{\textrm{trapped}}_{\nu}}.
    \label{eq:disttran}
\end{equation}

We consider the following neutrino production processes:
\begin{itemize}
\item $e^{-}+e^{+}\! \to\! \nu+\bar{\nu}$ (pair annihilation).
\item $p + e^{-}\! \to n + \nu_{e} \textrm{ or } n + e^{+}\! \to p + \bar{\nu}_{e}$ ($e^{-}$ and $e^{+}$ capture by nucleons).
\item $p + e^{-}\! \to n + \nu_{e}, \;  n + e^{+}\! \to p + \bar{\nu}_{e}$ (electron capture).
\item $A + e^{-}\! \to A' + \nu_{e}$ (plasmon decay).
\item $\tilde{\gamma}\!\to\nu+\bar{\nu}$ (nucleon-nucleon bremsstrahlung).
\end{itemize}

The emissivities and cross-sections for each process appear in~\cite{Dicus:1972yr,1975ApJ...201..467T,bruenn1985stellar,1996A&A...311..532R,2001PhR...354....1Y,Burrows2004,BURROWS2006356}. As in Section~\ref{subsec:spherical}, we use the two-flavor approximation with $\nu_{e},(\bar{\nu}_e)$ and $\nu_{x}(\bar{\nu}_{x})$, where $x$ represents a superposition of non-electron (anti)neutrinos.

The chemical equilibrium for electron or positron capture relates the neutrino degeneracy parameter $\eta_\nu$ to $\eta_{e}$ and $Y_e$, i.e.,
\begin{equation}
\eta_{e} - \eta_{\nu} =  \ln\left(\frac{1-Y_e}{Y_e}\right) + \frac{m_n-m_p}{k_B T}.
\label{eq:neutrinoschempot}
\end{equation}
\begin{figure*}
\centering
\includegraphics[width=0.49\hsize,clip]{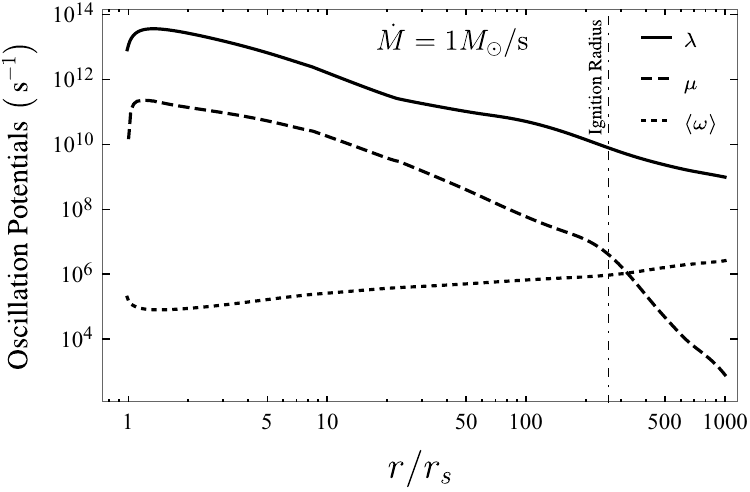}\includegraphics[width=0.49\hsize,clip]{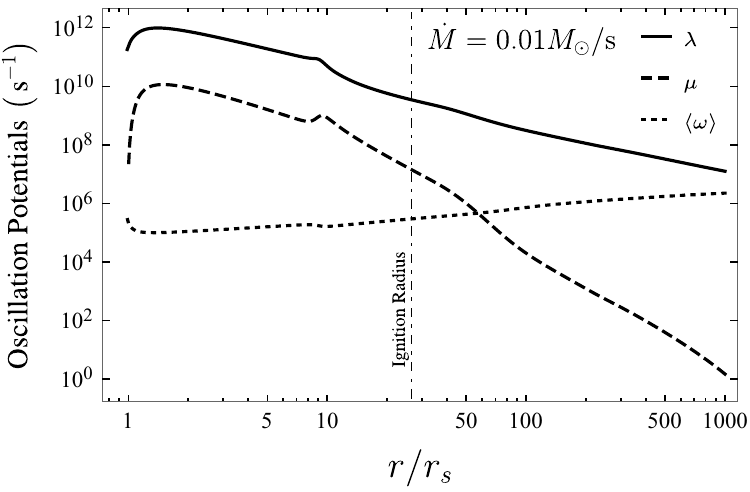}\\
\includegraphics[width=0.49\hsize,clip]{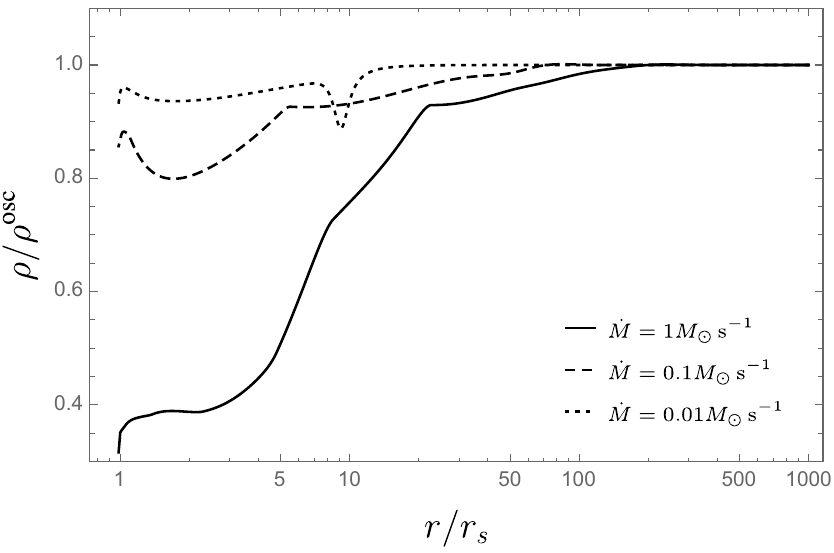}\includegraphics[width=0.49\hsize,clip]{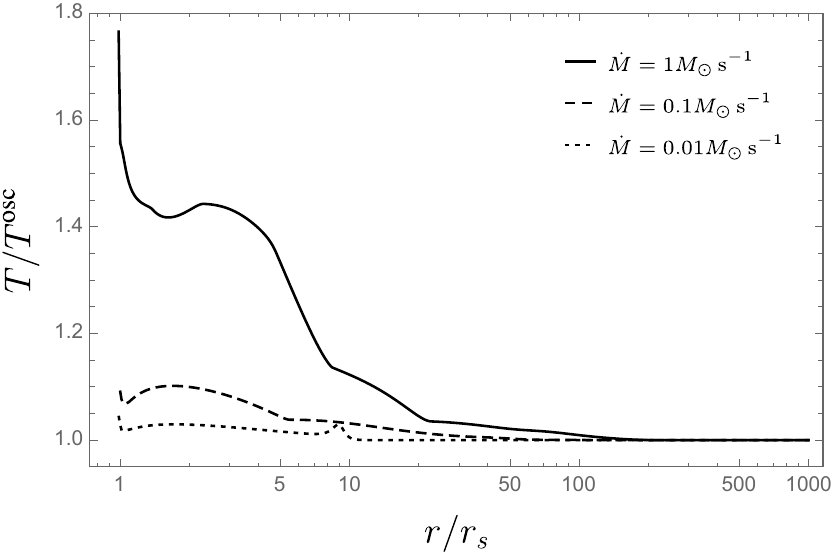}
\caption{Top panels: Oscillation potentials along the accretion disk for selected values of the accretion rate. The disk parameters are $M=3M_{\odot}$, $a = 0.95$ and $\alpha = 0.01$. Bottom panels: Ratio between the densities $\rho^{\rm osc}$ and $\rho$ (left panel) and the temperatures $T^{\rm osc}$ and $T$ (right panel) for a disk with flavor equipartition and a standard disk. Reproduced from~\cite{universe7010007}.} 
\label{fig:Potentials1}
\end{figure*}

We can use Eqs.~(\ref{eq:energycon}), (\ref{eq:leptonconservation}), (\ref{eq:press}) and (\ref{eq:neutrinoschempot}) to get the solutions $Y_{e},T,\eta_{e}$ as functions of the radial coordinate $r$, with the accretion rate, the viscosity and the BH spin ($\dot{M},\alpha,a$) as input parameters. \citet{2007ApJ...657..383C} and \citet{ LIU20171} estimated that neutrino cooling is efficient in NCADs for accretion rates in the range
\begin{equation}
\dot{M} \approx (0.01\text{--}1)\;M_{\odot}~\text{s}^{-1}.
\label{eq:mdotrange}
\end{equation}
Lower accretion rates lead to low neutrino production rates, and higher accretion rates trap the neutrinos within the disk. Accretion rates in Eq.~(\ref{eq:mdotrange}) are consistent with the BdHN scenario~\cite{2014ApJ...793L..36F, 2016ApJ...833..107B, 2019ApJ...871...14B, 2022PhRvD.106h3002B}.

It is worth mentioning that the NCAD model presents a degeneracy in the input parameters. A low value of $\dot{M}$, a low value of $a$, or a high value of $\alpha$ produces cold-dilute disks and vice versa~\cite {universe7010007}. We overcome this problem and decrease the number of free parameters by fixing the viscosity, the BH mass, and the spin at $M = 3 M_{\odot}$, $\alpha = 0.01$, and $a=0.95$, respectively. We chose to vary the accretion rate with these fixed parameters to compare and contrast our results with other disk models~(e.g., \cite{2007ApJ...657..383C}).

To solve Eqs.~(\ref{eq:energycon}) and (\ref{eq:leptonconservation}), we must set two conditions at the disk's external boundary. The NCAD model requires $U/\rho\sim M/r_{\rm ext}$, where $r_{\rm ext}$ is the circularization radius of the accreting matter~\cite{2015ApJ...812..100B,2016ApJ...833..107B}. The BdHN model suggests that the accreting matter contains predominantly oxygen and electrons, so we set $Y_{e}\vert_{r= r_{\rm ext}} = 1/2$.

In Fig.~\ref{fig:Disks}, we show $n_\nu$ and $E_{\nu}$ in the accretion disk. As in Section~\ref{subsec:spherical}, neutrino energies are of the order of MeV, and $n_{\nu_e}\sim n_{\bar{\nu}_e}$, while $n_{\nu_e}\gg n_{\nu_x}$. Yet, the two pictures are different. Since the disk has a physical extension determined by $H$ and $r$, one cannot set a symmetric two-dimensional emission surface. Also, curvature effects are important in the vicinity of the BH. However, the assumptions of the NCAD model simplify the analysis of oscillations. First, the vertically integrated, axially symmetric model implies that thermodynamic quantities are constant along the $z$ and $\phi$ directions. Second, defining a typical distance $\Delta r_{\rho,\text{eff}} = \left\vert d\ln\left( Y_{e}n_{B} \right)/dr\right\vert^{-1}$, we find $\Delta r_{\rho,\text{eff}} < 5$~$r_{s}$, which implies that we can consider the electron and neutrino densities as constants in thin neighborhoods inside the disk. Finally, we consider that an observer in the disk's comoving frame describes as isotropic gases the neutrino and lepton content in a small neighborhood at any point in the disk. In this frame, the equations of flavor evolution acquire a flat spacetime form and using the solutions in Fig.~\ref{fig:Disks}, we obtain the potentials inside the ignition radius (see top panels in Figure~\ref{fig:Potentials1}). When we solve the flavor evolution with these potentials, the oscillation length is
\begin{equation}
t_{\rm osc} \sim 1\mu{\rm s}.
    \label{eq:timeosc2}
\end{equation}

In symmetric $\nu\bar{\nu}$ gases (see Fig.~\ref{fig:Disks}), an anisotropic perturbation leads to kinematic decoherence and flavor equipartition~\cite{Raffelt:2007yz, EstebanPretel:2007ec}. The transition to equipartition lasts a few oscillation cycles in both mass hierarchies. In our case, the flux term, caused by the increase in density in the radial direction, leads to a steady-state disk with flavor equipartition. Introducing the condition
\begin{equation}
     P_{\nu_e \to \nu_e} =  P_{\bar{\nu}_e \to \bar{\nu}_e} = 1/2,
\end{equation}\label{eq:equipartition}
in the disk's hydrodynamic equations, we obtain a new set of solutions $Y^{\rm osc}_{e},T^{\rm osc},\eta^{\rm osc}_{e}$, which account for the oscillation dynamics. The bottom panels of Figure~\ref{fig:Potentials1} compare the two sets of solutions, i.e., a disk with flavor equipartition and a standard disk. The main consequence of flavor equipartition is to increase the density and decrease the temperature inside the ignition radius. The effect grows with the accretion rate and is consistent with the fundamental physics of the disk's cooling. Low accretion rates produce dilute disks with low neutrino optical depth. Neutrinos of all flavors can escape the disk, and equipartition has little to no effect on neutrino cooling, preserving (approximately) the values of thermodynamic variables but changing the outgoing flavor content. Conversely, high accretion rates produce dense disks. However, the optical depth of the electron neutrino is higher than the others ($\tau_{\nu_e} \gg \tau_{\bar{\nu}_{e}} \sim \tau_{\nu_{x}(\bar{\nu}_x)}$), inhibiting $\nu_e$ cooling. Flavor equipartition turns a large portion of electron neutrinos into non-electron neutrinos, increasing the efficiency of neutrino cooling and reducing the temperature inside the disk. A low temperature implies a low electron fraction and a high baryon density. Specifically, the ratio between the neutrino cooling for a disk with flavor equipartition and a standard disk obeys
\begin{equation}
\frac{F^{\text{osc}}_{\nu}}{F_{\nu}} = \frac{1}{2}\left(1 + \frac{\langle E_{\nu_{x}} \rangle}{\langle E_{\nu_{e}} \rangle}\frac{1 + \tau_{\nu_{e}}}{1 + \tau_{\nu_{x}}}\right). 
    \label{eq:fluxcomp}
\end{equation}

The change in the flavor content emitted by the disk (fewer electron neutrinos) decreases the energy density of the $e^+e^-$ plasma generated by $\nu\bar{\nu}$ annihilation. We refer to \cite{universe7010007} for further details.
%

%%%%%%%%%%%%%%%%%%%%%%%%%%%%%%%%%%%%%%%%%%%%%%%%%%%%%
%%%%%%%%%%%%%%%%%%%%%%%%%%%%%%%%%%%%%%%%%%%%%%%%%%%%%
\section{GeV-TeV neutrinos from BdHN I}
\label{sec:4}
%%%%%%%%%%%%%%%%%%%%%%%%%%%%%%%%%%%%%%%%%%%%%%%%%%%%%%
%%%%%%%%%%%%%%%%%%%%%%%%%%%%%%%%%%%%%%%%%%%%%%%%%%%%%

We turn now to the production mechanism of neutrinos of higher energy, in the GeV-TeV domain, in the physical setup of BdHN I. As we mentioned in Section \ref{sec:1}, the $pp$ interactions of the protons engulfed by the expanding $e^+e^-$ plasma (created via vacuum polarization around the newborn BH) with the protons ahead of it produce those neutrinos.

To set up the possible $pp$ interactions occurring in a BdHN I, we start by analyzing the structure of the baryonic matter present. For this task, we use three-dimensional simulations of this system~ (see, e.g., \cite{2016ApJ...833..107B, 2019ApJ...871...14B, 2022PhRvD.106h3002B}). Although the SN ejecta starts expanding with spherical symmetry, it becomes asymmetric by the accretion process onto the NS and the BH formation~\cite{2019ApJ...883..191R}. Due to this morphology, the $e^+e^-$ plasma created in BH formation, which expands isotropically from the newborn BH site, experiences different dynamics along different directions due to the different amounts of baryonic matter encountered~\cite{2018ApJ...852...53R}.

In hydrodynamic simulations, the dynamics of the $e^+e^-$ plasma around the BH (see, e.g., \cite{1998A&A...338L..87P, 1999A&AS..138..511R, 2000A&A...359..855R, 2001A&A...368..377B}) depends on the number of baryons in the plasma set by the \textit{baryon load} parameter, which is the ratio of the rest-mass energy of the baryons (in the $e^+e^-$ plasma) to the $e^+e^-$ energy, i.e., $B\equiv M_b c^2/E_{e^+e^-}$. Along the line from the CO star to the accreting NS, the NS accretion process and the BH formation cave a region characterized by very poor baryon pollution, a \textit{cavity} (see Fig. \ref{fig:Interactionsscheme1}; also Refs.~\cite{2016ApJ...833..107B, 2019ApJ...871...14B, 2019ApJ...883..191R}). In this situation, $B\lesssim 10^{-2}$ leads to the plasma transparency with a high Lorentz factor $\Gamma\sim 1/B\gtrsim 10^2$ (see, e.g., \cite{2000A&A...359..855R}). We denote with $\gamma$ the Lorentz factor of a single particle, and with $\Gamma$ the one of bulk motion. In other directions along the orbital plane, the $e^+e^-$ plasma penetrates the SN ejecta at $\sim 10^8$--$10^{10}$~cm, so it engulfs much more baryons. Under these conditions, the plasma reaches transparency at $10^{12}$~cm with $\Gamma\lesssim 4$ (see Fig. \ref{fig:Interactionsscheme1}). The theoretical description and numerical simulations of the $e^+e^-$ plasma dynamics with large baryon load ($B\sim 100$) have been presented in Ref.~\cite{2018ApJ...852...53R}.

With the above physical and geometrical description in mind, we set up the properties of the incident and target protons for two types of $pp$ interactions in a BdHN I:
\begin{enumerate}
    \item
    Interaction of protons with $\Gamma \lesssim 6$ within the self-accelerated $e^+e^-p$ plasma that penetrates the SN ejecta, with the unshocked protons ahead the plasma expansion front, at rest inside the ejecta (see Fig.~\ref{fig:Interactionsscheme1}). This situation leads to $B\sim 50$ (see Fig.~\ref{fig:Interactionsscheme1} here and Figs. 34, 35, and 37 in \cite{2018ApJ...852...53R}).
    \item 
    Interaction of protons with $\Gamma \sim 10^2$--$10^3$ engulfed in the self-accelerated $e^+e^-p$ plasma in the direction of least baryon density around the newborn BH, with the protons at rest of the interstellar medium (ISM) (see Fig.~\ref{fig:Interactionsscheme1}). We assume interaction occurs at $\sim 10^{16}$--$10^{17}$~cm away from the system, as inferred from the time and $\Gamma$ at transparency (see, e.g., \cite{2012A&A...543A..10I}). This situation leads to $B\sim 10^{-2}$--$10^{-3}$ (see Fig.~\ref{fig:Interactionsscheme1} of here, and Figs. 34, 35 in \cite{2018ApJ...852...53R}).
\end{enumerate}

The above scenario of $pp$ interactions is markedly different from previous works based on the collapsar and fireball model with relativistic shocks (see, e.g., \cite{razzaque2004tev, ando2005revealing, razzaque2003neutrino}, for details). In the latter, collimated jets accelerate protons that produce very energetic neutrinos from several hundreds of GeV to several TeV. In the BdHN model, the plasma expands in all directions, with different amounts of engulfed matter from the SN ejecta, leading to secondary emerging particles of different energies. Thus, we do not deal with a collimated jet, protons are less energetic, and the neutrino energy is of a few GeV to TeV.

%%%%%%%%%%%%%%%%%%%%%%%%%%%%%%%%%%%%%%%%%%%%%%%%%%%%%%%%
\subsection{Neutrino production in the high-density ejecta}
\label{sec:4a}
%%%%%%%%%%%%%%%%%%%%%%%%%%%%%%%%%%%%%%%%%%%%%%%%%%%%%%%%

We analyze here the $pp$ interactions that occur when the $e^+e^-\gamma$ plasma starts to engulf the baryons present in the SN ejecta, forming an $e^+e^-\gamma p$ plasma (see Fig.~\ref{fig:Interactionsscheme1}). These accelerated baryons interact with the target baryons ahead of them (at rest), producing secondary pions that subsequently decay as $ \pi\rightarrow \mu \nu_{\mu}$, $\mu\rightarrow e+\nu_e+\nu_\mu$ (for charged pions) and $\pi^0\rightarrow 2\gamma$ (for neutral pions). We perform relativistic hydrodynamic (RHD) simulations of such an $e^+e^-$ plasma. We describe the RHD simulations in the following subsection, which allows us to estimate the Lorentz factor and the energy of the incident and target photons participating in the interactions.

%
%%%%%%%%%%%%%%%%%%%%%%%%%%%%%%%%%%%%%%%%%%%%%%%%%%%%%%%%
\subsubsection{RHD simulations}
\label{S:RHDsimulations}
%%%%%%%%%%%%%%%%%%%%%%%%%%%%%%%%%%%%%%%%%%%%%%%%%%%%%%%%
%
Here, we summarize the equations that govern the $e^+e^-$ plasma expansion inside the SN ejecta within the BdHN scenario. We have run RHD simulations (see Ref.~\cite{2018ApJ...852...53R} for additional details) performed with a one-dimensional implementation of the RHD module of the \texttt{PLUTO} code \cite{mignone2011pluto}. 
The code integrates a system of partial differential equations in distance and time, which allows computing the evolution of the thermodynamical variables and dynamics of the $e^+e^-$ plasma. The plasma carries baryons from the ejecta, along one selected radial direction, at any time. The equations are those of ideal relativistic hydrodynamics (see Section~10 in Ref.~\cite{2018ApJ...852...53R}).

The plasma equation of state is that of a relativistic ideal gas with an adiabatic index of $4/3$ (see Appendix B in Ref.~\cite{2018ApJ...852...53R}). The simulation starts at the moment of BH formation, so we set the initial conditions from the final configuration of the numerical simulations in Ref.~\cite{2016ApJ...833..107B}:
\begin{enumerate}
    \item 
    The CO star has a total mass of $11.15~M_\odot$ distributed as $2 M_\odot$ of the $\nu$NS and $9.15 M_\odot$ of ejecta mass (envelope mass). At the SN explosion time, the ejecta profile follows a power-law profile $\rho \propto r^{-2.8}$ (see, e.g., Ref.~\cite{2016ApJ...833..107B}).
    \item 
    The orbital period and binary separation are $P\approx 5$~min and $a\approx 1.5\times 10^{10}$~cm.
    \item
    The pressure and velocity of the ejecta are negligible with respect to the corresponding properties of the plasma. Therefore, we consider the remnant at rest as seen from the plasma.
    \item
    The baryon load of the $e^+e^-$ plasma is not isotropic since the density is different along different directions. According to the three-dimensional simulations of Ref.~\cite{2016ApJ...833..107B}, the ejecta density profile along a given direction, at the BH formation time, decays with distance as a power-law, i.e., $\rho \propto (R_0-r)^\alpha$ (see, e.g., Figs.~34--35 of Ref.~\cite{2018ApJ...852...53R} that show the mass profiles along selected directions). The normalization, the constant $R_0$, and the parameter $2<\alpha<3$ depend on the angle.
    \item 
    The total isotropic energy of the $e^+e^-$ plasma is set to $E_{e^+e^-}=3.16\times 10^{53}$~erg. Therefore, the baryon load parameter is $B=51.75$ in the high-density region.
\end{enumerate}
The evolution from these initial conditions leads to the formation of a shock and its subsequent expansion until the outermost ejecta regions. The relevant radial distances in the simulation are $\sim10^{9}$--$10^{10}$~cm.

Throughout the expansion, the $e^+e^-$ plasma continuously \textit{phagocytoses} baryons, so the spectrum of the secondary particles, the proton energy distribution, and the baryon number density depend on the radial position of the shock. By taking snapshots of this process, we obtain the relative spectrum for each secondary particle within a thin shell close to the shock. We integrate all these spectra over the radius to estimate the energy released through the different channels.

Considering that the protons follow a Maxwell-Boltzmann energy distribution in the comoving frame, the energy distribution in the laboratory frame is peaked enough to be well approximated by a delta function. Hence, we consider a monochromatic proton energy distribution $J_p(E_p)$, i.e., $J_p(E_p)\propto \delta(E_p-E_p^0)$. The value of $E_p^0$ depends on the Lorentz factor $\gamma(r)$. Due to momentum-energy conservation, $\gamma$ decreases rapidly with time. Therefore, we focus on the first stages of the expansion when protons have enough energy to interact. We estimate the interactions from a radius $r_i$ where the $\gamma e^\pm$ plasma with engulfed protons has the maximum Lorentz factor, up to a final radius, $r_f$, over which the proton energy becomes lower than the interaction threshold energy. From our numerical simulation, we find that this region extends from $r_i \approx 9.6\times 10^8$~cm to $r_f \approx 3.0\times 10^{10}$~cm, so $\Delta r=r_f-r_i \approx 3\times 10^{10}$~cm. This thickness is much smaller than the extension of the SN ejecta, which is of the order of $ 10^{12}$~cm~\cite{2018ApJ...852...53R}.

Next, we describe how we extract from these simulations the physical quantities that we use to compute the particles spectra, i.e., the protons Lorentz factor, the number density of the incident and target protons, each of them considered at every radius of the expansion of the shock inside the ejecta.

%%%%%%%%%%%%%%%%%%%%%%%%%%%%%%%%%%%%%%%%%%%%%%%%%%%%%%%%
\subsubsection{Physical quantities for the \textit{pp} interaction}
\label{S:PhysicalQuantities}
%%%%%%%%%%%%%%%%%%%%%%%%%%%%%%%%%%%%%%%%%%%%%%%%%%%%%%%%

The baryons of the SN ejecta are incorporated time by time, at every radius, by the $e^{\pm}\gamma$ plasma. The incident protons are those engulfed by the expanding shock front. The target protons are those ahead of the shock front, which we assume at rest regarding the incident protons. Having clarified this, we identify the physical quantities needed to calculate the spectra at each radius: the Lorentz factor of the protons in the shock front, $\gamma_p$, their energy, their density, $n_{\rm sh}$, and density of the unshocked protons, $n_t$. These quantities change at every radius as the plasma expands. We calculate all these quantities in the laboratory frame.

We compute the above quantities as follows. First, we obtain the position of the shock front from the simulation, i.e., $r_{\rm front}$, given by the abrupt fall of the pressure in the ejecta. At this radius, the protons in the shock have a maximum Lorentz factor. Although the pressure at $r>r_{\rm front}$ falls fast, the extension of this region is smaller than the mean free path of the front protons, defined by $\lambda_p^{-1}=\sigma_{\rm pp}\left(E_p\right)\times n_p$, where $\sigma_{\rm pp}$ is the $pp$ cross-section and $n_p$ their number density. Thus, to calculate the density of incident and target protons, we average all the possible interacting protons at a given time. For the incident protons density, $\langle n_{\rm sh}\rangle$, we average the radial density in the region $r_{\rm front}-\lambda_p<r< r_{\rm front}$, e.g., the incident density at an average radius $\langle r_{\rm front}\rangle$. A similar average applies ahead of the front, i.e., at $r_{\rm front}<r<r_{\rm front}+\lambda_p$, to obtain the density of the target protons, $\langle n_{\rm t}\rangle$. 

\begin{figure}
\includegraphics[width=\hsize,clip]{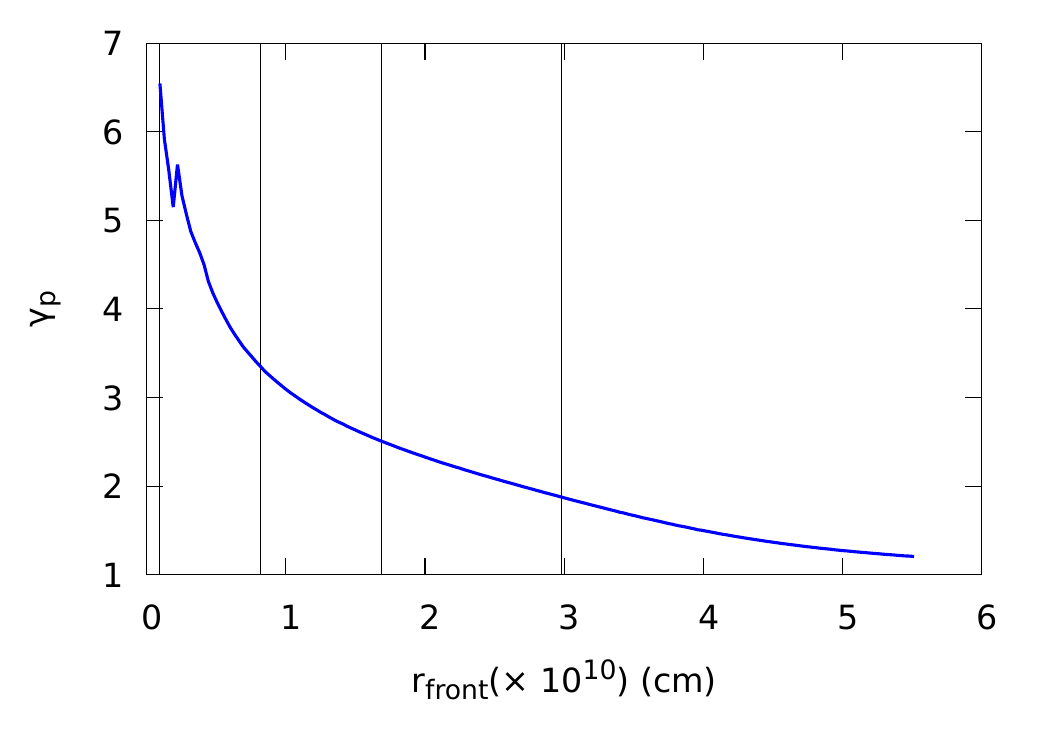}
\caption{Evolution of the Lorentz factor of the protons in the shell front, $\gamma_p$, as a function of the (average) front radius position. We recall that $\gamma_p=\Gamma(r_{\rm front})$, since outside the front the protons are roughly at rest with respect to the shell (the velocity of the remnant is much slower). The vertical lines are four selected radii: $r_1=r_i=9.59\times 10^{8}$~cm, $r_2=8.19 \times 10^9$~cm, $r_3=1.69\times 10^{10}$~cm, and $r_f=2.98\times 10^{10}$~cm. At position $r=r_i$, the protons have the maximum $\gamma$ factor.}
\label{fig:fig3}
\end{figure}

Then, we calculate the maximum value of $\gamma_p$ inside the shell and, correspondingly, the energy of the protons, $E_p(r)=\gamma_p(r) m_p c^2$. The proton Lorentz factor $\gamma^p_k$, at the generic radius $r_k$, is given by the value of the baryons velocity, $\beta^p_k$, at the shock front position $\langle r_{\rm front}\rangle$. We show in Fig.~\ref{fig:fig3} the profile of the maximum values of the Lorentz factor at every front radius. The numerical simulations of the expansion of the plasma inside the ejecta give us a particle velocity distribution (see Fig. 37 in \cite{2018ApJ...852...53R}). From the distribution, we extract the maximum Lorentz factor consistent with the density average process explained above. We note that the proton Lorentz factor in Fig.~\ref{fig:fig3} corresponds to the one of the shell bulk motion $\Gamma$, that is $\gamma_p=\Gamma(r_{\rm front})$. The correspondence is justified mainly by two circumstances. First, the bulk motion accelerates the particles it engulfs at each radius. Second, the region under consideration is relatively small (see Fig.~\ref{fig:fig3} and below), which corresponds to an expansion time interval of $\sim 1$~s (see Fig.~\ref{fig:fig11}). Consequently, the self-acceleration of the plasma and its bulk motion accelerate and drive the motion of the engulfed particles at any time.

\begin{figure}
\includegraphics[width=\hsize,clip]{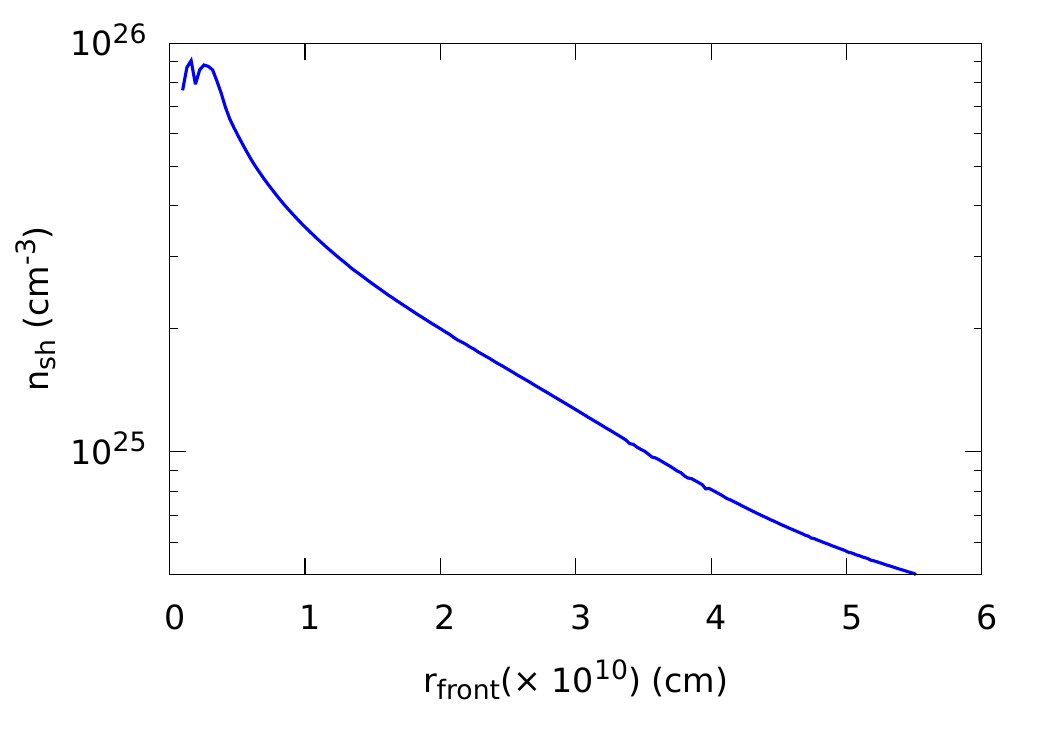}
\caption{Average baryons number density of the expanding shell at the front position  $r_{\rm front}$.}
\label{fig:fig4}
\end{figure}

From the above, the energy of protons is $1.24 \leq E_p\leq 6.14$~GeV, which is enough to produce secondary particles. The proton energy threshold for pion production is, for the interaction $pp\rightarrow p n\pi^+$, $E_{p,\rm Th}=1228$~MeV and, for $pp\rightarrow pp \pi^0$, $E_{p,\rm Th}=1217$~MeV. The protons with the highest energy ($\gamma\sim 6$) dominate the neutrino production at these low energies. Figure~\ref{fig:fig3} shows four vertical lines at fixed front radii of reference: the first vertical line corresponds to $r_i$ where the protons have their maximum energy, while the last line corresponds to $r_f$ (there, $\gamma_p \approx1.878$). The intermediate radii show the evolution of the particle spectrum during the expansion. We now compute the particle spectra at these four specific radii. At every radius, the average number density of the target protons $\langle n_t \rangle$ in the remnant varies between $8\times 10^{23}$~cm$^{-3}$ at $r_i$ to $\sim 5\times 10^{23}$~cm$^{-3}$ at $r_{\rm end}=5.51\times 10^{10}$~cm (the endpoint of the simulation). The protons number density at the front of the expanding shell, $\langle n_{\rm sh} \rangle$, does not vary much; it is in the range $(0.5$--$9)\times10^{25}$~cm$^{-3}$. The maximum value occurs in the region close to the initial radius $r_i$, and the lower value to the final radius $r_{\rm end}$, as shown in Fig.~\ref{fig:fig4} that plots the density as a function of the front radius, consistently with Fig.~\ref{fig:fig3}. 

%%%%%%%%%%%%%%%%%%%%%%%%%%%%%%%%%%%%%%%%%%%%%%%%%%%%%%%%%
\subsubsection{Particles spectra}\label{S:ParticlesSpectra}
%%%%%%%%%%%%%%%%%%%%%%%%%%%%%%%%%%%%%%%%%%%%%%%%%%%%%%%%%

We now turn to the spectra of the emerging particles from the $\pi$ and $\mu$ decay. For the calculation of the pion production rate, we use the parameterization for the pion production cross-section presented in Ref.~\cite{blattnig2000parametrizations}. They provide a formula for the production of the three types of pions ($\pi^0,\pi^+,\pi^-$) as a function of the pion and incident proton energy, $d\sigma(E_{\pi}, E_p)/dE_{\pi}$, in two ranges of incident protons kinetic energy in the laboratory frame $0.3\leq T_p^{\rm lab}\leq 2$~GeV and $2\leq T_p^{\rm lab}\leq 50$~GeV. This parameterization of the cross-section is appropriate for our calculations since it is accurate in the energy regime of interest, namely $E_p< 7$ GeV. Thus, the pion production rate can be computed as
\begin{equation}\label{2}
Q_{\pi}(E_{\pi})=c n_p \int_{E_{\pi}}^{E_p^{\rm max}} J_p(E_p) \frac{d \sigma}{dE_{\pi}} dE_p,
\end{equation}
where $\sigma = \sigma(E_{\pi},E_p)$, $J_p(E_p)$ is the proton energy distribution, $n_p$ the number density of target protons, $c$ the speed of light, and $E_p^{\rm max}$ the maximum proton energy in the system. Since we consider a fixed value for the proton energy, $E_p^0$, at the front of each spherical shell, we assume $J_p(E_p)=A \delta(E_p-E_p^0)$, where $A$ is the baryon number density at the shell front. Thus, the production rate $Q_{\pi}$ becomes
\begin{equation}
    \label{3}
    Q_{\pi}\left(E_{\pi}\right)=c n_p A \frac{d\sigma}{d E_{\pi}} \theta\left(E_p^0-E_{\pi}\right) \theta\left(E_p^{\rm max}-E_p^0\right),
\end{equation}
where now $\sigma = \sigma\left(E_{\pi},E_p^0\right)$. With Eq.~(\ref{3}) for the $\pi$ production rate, we can compute the spectra for all the particles. Because the cross-section for neutral, negative, and positive pions are different, we need to distinguish between emerging particles from $\pi^0$ decay in two photons, $\pi^-$ decay: $\pi^-\rightarrow \mu^-\,\bar{\nu}_{\mu^{(1)}};\,\mu^-\rightarrow e^-+\bar{\nu}_e+\nu_{\mu^{(2)}}$ and from $\pi^+$ decay: $\pi^+\rightarrow \mu^+\,\nu_{\mu^{(1)}};\,\mu^+\rightarrow e^++\nu_e+\bar{\nu}_{\mu^{(2)}}$ decay. 

We denote the spectrum of the produced particle $a$ as $\Phi_a = dN_a/dE_a$, where we indicate with $N_a$ the particle number density per unit of time. We denote as $\nu_{\mu^{(1)}}$ the muon neutrino/antineutrino from the direct pion decay, $\pi\rightarrow \mu \nu_{\mu}$, and $\nu_{\mu^{(2)}}$ the neutrino/antineutrino from the consequent muon decay, $\mu\rightarrow e \nu_{\mu} \nu_e$.

The spectrum of photons from $\pi^0$ decay is given by
\begin{equation}
\label{4}
\Phi_{\pi^0\rightarrow\gamma\gamma}\left(E_{\gamma}\right)=2\int_{E^{\rm min}\left( E_{\gamma}\right)}^{E_{\pi}^{\rm max}} \frac{Q_{\pi}(E_{\pi})}{\sqrt{E_{\pi}^2-m_{\pi}^2c^4}} dE_{\pi},
\end{equation}
where $E^{\rm in}\left(E_{\gamma}\right)=E_{\gamma}+m_{\pi^0}^2c^4/(4 E_{\gamma})$ can be derived by the kinematics of the process. The factor $2$ takes into account the two produced photons, while $Q_{\pi}\left(E_{\pi}\right)$ is given by Eq.~\eqref{3}, with the corresponding pion spectral distribution for $\pi^0$ (see Ref.~\cite{blattnig2000parametrizations}).

\begin{figure}
\centering
\includegraphics[width=\hsize,clip]{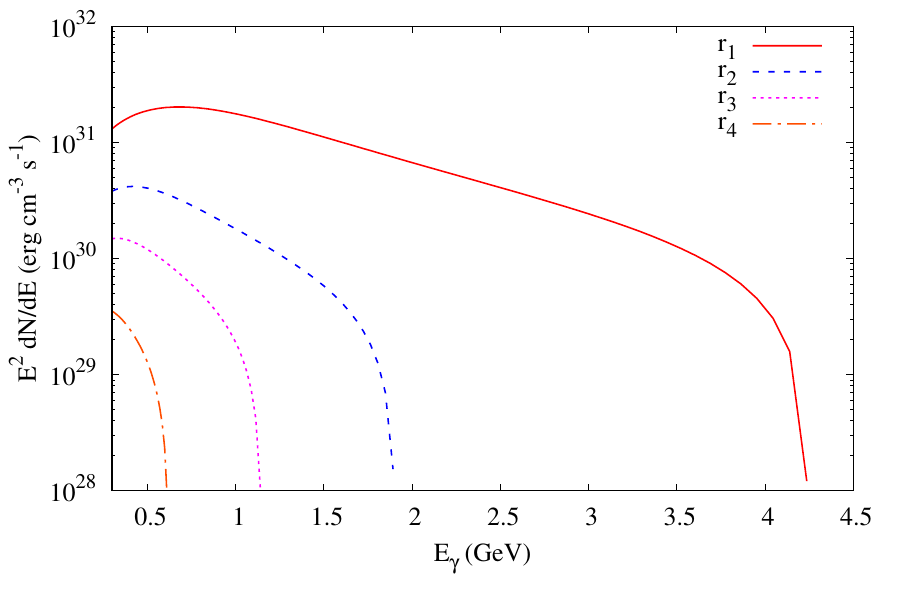}
\caption{Photons spectrum from the decay $\pi^0\rightarrow \gamma\gamma$. The profiles are shown at four selected radii for the expansion of the shell inside the ejecta: the radii $r_{2,3,4}$ are the same as Fig.~\ref{fig:fig3}, while $r_1=1.28\times 10^9$~cm. At $r\gtrsim r_4$, the proton energy approaches (from higher energies) the interaction threshold energy producing an emission cutoff.}
\label{fig:fig5}
\end{figure}

We show the photon emissivity in Fig.~\ref{fig:fig5}. The total energy, integrated over all photons energies and calculated via Eq.~(\ref{eq:etotintegV}) (see later in Section~\ref{2C}), in the emissivity region, is given in Table~\ref{tab:1}.

\begin{table}
\centering
\begin{tabular}{*{3}{c}}
\hline
        Particle                & Total energy  \\
                                & ($10^{51}$~erg)  \\
\hline
$\gamma$ & $44.62$\\                                
\hline
\hline
$\nu_{\mu^{(1)}}$; $\bar{\nu}_{\mu^{(1)}}$  & $0.471$; $0.155$ \\
\hline
Without polarization & \\
\hline
$\nu_{{\mu}^{(2)}}$; $\bar{\nu}_{\mu^{(2)}}$ & $0.603$; $3.534$ \\
\hline
$\nu_e $; $\bar{\nu}_e$ &  $2.105$; $0.369$ \\
\hline
\hline
With polarization & \\
\hline
$\nu_{{\mu}^{(2)}}$; $\bar{\nu}_{\mu^{(2)}}$ & $2.307$; $2.854$ \\
\hline
$\nu_e $;  $\bar{\nu}_e$ & $2.825$ ; $0.494$ \\
\hline\end{tabular}
\caption{Total energy, integrated over the whole emitting region, via  Eqs.~(\ref{eq:etotintegV}), for $\gamma$,$~\nu_{\mu^{(1)}},~\nu_{\mu^{(2)}}$ and $\nu_e$, within and without considering the polarization (only for the last two particles). If we sum all the energies for all the considered $\nu$s, besides the $\nu_{\mu^{(1)}}+\bar{\nu}_{\mu^{(1)}}$, only for the case without or whit polarization (we consider the case with polarization), we obtain a total energy release of $9.11\times 10^{51}$~erg, that is $\approx 2.9\%$ of the energy of initial energy of the $\gamma e^{\pm}$ plasma. If we also include the energy emitted in photons, we get a total energy of $5.37\times 10^{52}$~erg, which corresponds to $17\%$ of the isotropic energy of the plasma, $E_{e^+e^-}$.}
\label{tab:1}
\end{table}

The neutrino spectrum from direct pion decay $\pi\rightarrow \mu\nu_{\mu}$ is given by
\begin{equation}\label{5}
\Phi_{\pi\rightarrow\mu\nu_{\mu}}\left(E_{\nu_{\mu}}\right)=\frac{1}{\lambda}\int_{E^{\rm min}\left( E_{\nu_{\mu}}\right)}^{E_{\pi}^{\rm max}} \frac{Q_{\pi}(E_{\pi}) \theta\left(\lambda-\frac{E_{\nu_\mu}}{E_{\pi}}\right)}{\sqrt{E_{\pi}^2-m_{\pi}^2c^4}} dE_{\pi},
\end{equation}
where $E_{\pi}^{\rm max}$ and $E^{\rm min}\left(E_{\nu_{\mu}}\right)$ are derived from the process kinematics. The lower limit of the integral is $E^{\rm min}\left(E_{\nu_{\mu}}\right)=E_{\nu_{\mu}}/\lambda+\left(\lambda/4\right) \left(m_{\pi}^2c^4/E_{\nu_{\mu}}\right)$, where $\lambda=1-r_\pi$ and $r_\pi=(m_\mu/m_\pi)^2$, being $r_\pi$ the maximum energy fraction that the neutrino emerging from the direct decay can take from the pion. The upper limit of the integral, $E_{\pi}^{\rm max}(E_p)$, can be derived by calculating the pion energy in the center-of-mass frame.

\begin{figure}
\includegraphics[width=\hsize,clip]{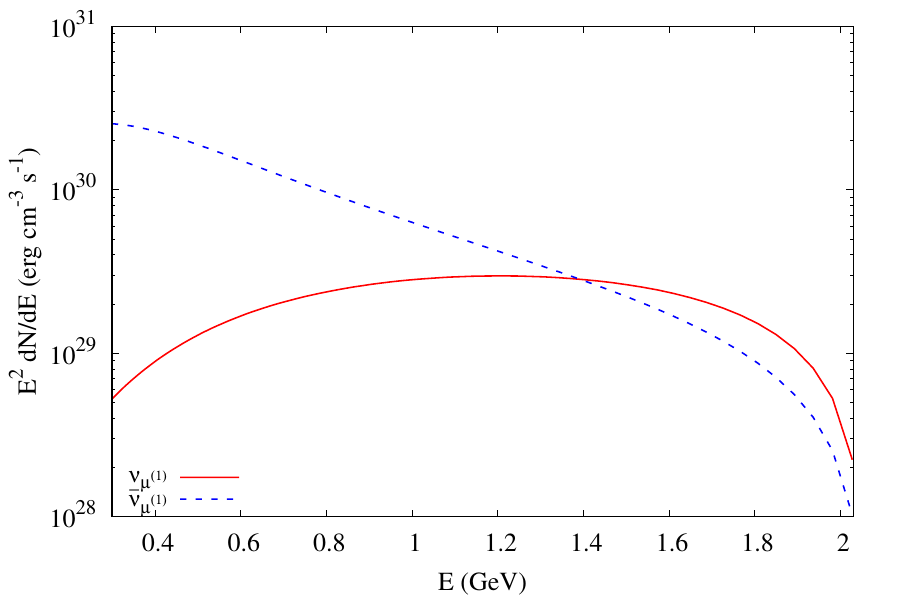}
\caption{Spectra of direct muon neutrinos from $\pi$-decay, at the radius $r_{1}$ of Fig.~\ref{fig:fig3}.}
\label{Numu1SpectraHighDensity}
\end{figure}

The spectra derived by Eq.~(\ref{5}) must be calculated via Eq.~(\ref{2}) using the parameterization of the cross-section for $\pi^-:\,d\sigma_{\pi^-}(E_{\pi};E_p)/dE_{\pi}$, and for $\pi^+:\,d\sigma_{\pi^+}(E_{\pi};E_p)/dE_{\pi}$, given in \cite{blattnig2000parametrizations}. The $\nu_{\mu^{(1)}}$ emissivities at radius $r_1$ are shown in Fig.~\ref{Numu1SpectraHighDensity}. The total energy, integrated over the region of emissivity, is given in Table~\ref{tab:1}.

The neutrino spectra from the decays $\pi\rightarrow \mu\rightarrow\nu$ can be calculated as
\begin{equation}
\label{6}
\Phi_{\pi\rightarrow\mu\rightarrow\nu}\left(E_{\nu}\right)=\int_{E^{\rm min}\left(E_{\nu}\right)}^{E_{\pi}^{\rm max}} \frac{Q_{\pi}(E_{\pi})}{\sqrt{E_{\pi}^2-m_{\pi}^2c^4}}\,g(z) dE_{\pi},
\end{equation}
where the pion production rate $Q_{\pi}$ is given in Eq.~(\ref{3}), and $z=E_{\nu}/E_{\pi}$. The functions $g(z)$ represent the $\nu$ spectra after the decay chain. We use the ultrarelativistic limit ($\beta_{\pi}\rightarrow 1,\,\beta_{\mu}\rightarrow 1$) of the Appendix A3 of \cite{LIPARI1993195}.

The function $g(z)$ can be decomposed as the sum of an unpolarized spectrum, $g^0(z)$, plus a polarized one, $g^{\rm pol}(z)$, $g(z)=g^0(z)+g^{\rm pol}(z)$. The functions $G(z)$, polarized and unpolarized, can be found in the Appendix of Ref.~\cite{LIPARI1993195}. Here, the limit $\beta_{\pi}\rightarrow 1$ is satisfied. Indeed, from the kinematics, we obtain that the pion Lorentz factor lies in the range $4.5\leq \gamma_{\pi}\leq 34.5$.

To have an expression for the spectrum of the particles coming from the $\mu$-decay, we have to insert Eq.~(\ref{3}) into Eq.~(\ref{6}). The $G(z)$ equations for $\nu_{\mu^{(2)}}$ and for $\nu_e$ (for unpolarized muon) are given in Eqs.~(102)-(103) in the Appendix A.3 of \cite{LIPARI1993195}.

\begin{figure}
    \centering
    \includegraphics[width=\hsize,clip]{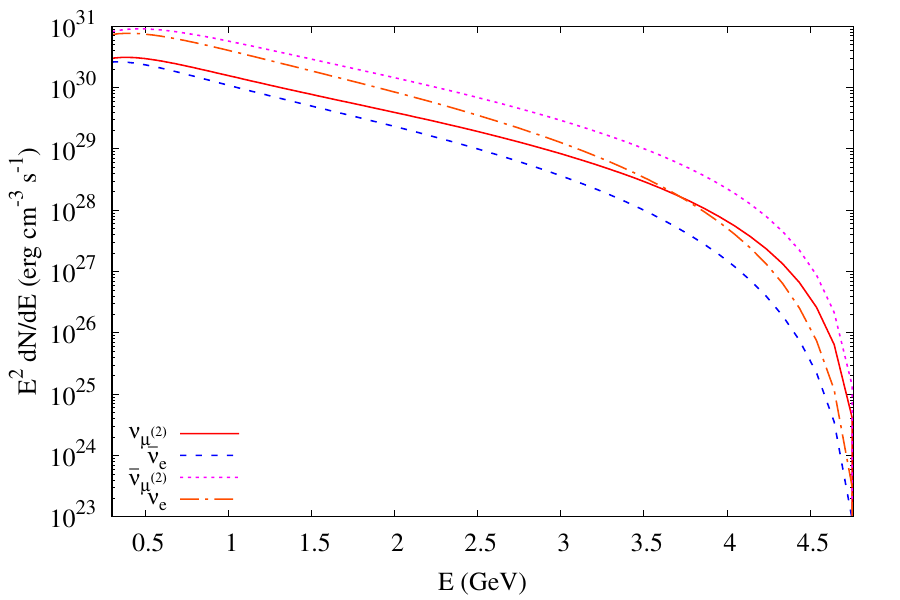}
    \caption{Spectra of muon and electron neutrinos emerging from $\mu^{\pm}$-decay, at the radius $r_1$.}
    \label{Numu2NueSpectraHighDensity}
\end{figure}

The minimum integration value $E^{\rm min}$ derive from the kinematic and is the same for $\nu_{\mu}$ and $\nu_e$, $E^{\rm min}\left(E_{\nu}\right)=E_{\nu}+ m_{\mu}^2c^4/(4E_{\nu})$. The emissivities for $\nu_{\mu^{(2)}}/\bar{\nu}_{\mu^{(2)}}$ and $\nu_e/\bar{\nu}_e$ are shown in Fig.~\ref{Numu2NueSpectraHighDensity}. 

We have also considered the effect of polarization, adding the formula for the polarized spectrum given in Eqs.~(104)--(105) in~\cite{LIPARI1993195}. However, we did not find substantial differences with respect to the unpolarized spectrum. The small quantitative difference in the luminosity is shown in Fig.~\ref{fig:fig11} and Table \ref{tab:1}. 

%%%%%%%%%%%%%%%%%%%%%%%%%%%%%%%%%%%%%%%%%%%%%%%%%%%%%%%%%%
\subsubsection{Total luminosity and total energy release}\label{2C}
%%%%%%%%%%%%%%%%%%%%%%%%%%%%%%%%%%%%%%%%%%%%%%%%%%%%%%%%%%

As we have seen from the above formulation, we can obtain the particle spectra at every radius $r_i$, which we denote hereafter as $\Phi^i_a(E_a)$. Thus, the particle emissivity at every radius, $\epsilon^i_a$, is given by
\begin{equation}
    \label{12}
    \epsilon^i_a=\int_{0.3\,\rm GeV}^{E_{\pi}^{\rm max}} \Phi^i_a( E_a)\,E_a dE_a,
\end{equation}
where $E_{\pi}^{\rm max}$ is the maximum pion energy derived from the kinematic of the process.

Then, the power (``luminosity'') emitted in particles of type $a$, at the radius $r_i$, is
\begin{equation}\label{eq:luma}
    L^i_a = \int_{V_i} \epsilon^i_a dV,
\end{equation}
where the integration is carried out over the volume $V_i$ of the emitting/interacting shell at the front position $r_i$.

The total emissivity and luminosity at the radius $r_i$ are the sum of the contributions of all particles, i.e.,
\begin{equation}\label{eq:emlumtotal}
    \epsilon^i_{\rm tot}= \sum_{a}\epsilon^i_a,\qquad
     L^i_{\rm tot} = \sum_{a}L^i_a.
\end{equation}
The energy emitted in $a$-type particles is given by
\begin{equation}\label{eq:enparticle}
    {\cal E}_a=\int L^i_a (t)\,dt,
\end{equation}
where the integration is carried out along the duration of the emission. Therefore, the total energy emitted in all the emission processes is
\begin{equation}\label{eq:entot}
    {\cal E}=\sum_{a}{\cal E}_a.
\end{equation}

From the numerical simulation of the expanding shell inside the remnant, we know that the width of the shell is $\Delta r_{\rm sh}\approx 3\times10^8$~cm. Since the mean free path of the interaction is much smaller than the shell width, the interacting volume at the radius $i$ is approximately $V_i=4\,\pi\,r_i^2\,\lambda_i$, where $\lambda_i$ is the mean free path of the protons of energy $E^i_p$ in the shell front. The mean free path is given by $\lambda_i=\left(\sigma_{\pi^{\pm,0}} A \right)^{-1}$, where $A$ is the baryon number density at the front, and $\sigma_{\pi^{\pm,0}}$ is the inclusive cross-section for $\pi^-$, $\pi^{+}$ and $\pi^0$ (see Section~3 in \cite{blattnig2000parametrizations}). For $\pi^+$, $0.4\leq \lambda^i_{\pi^+}\leq 11$~cm; for $\pi^-$, $1.18 \leq \lambda^i_{\pi^-}\leq 45$~cm; for $\pi^0$, $0.65 \leq \lambda^i_{\pi^0}\leq 50.4$~cm. We calculate the mean free path for each pion at the initial and final radius shown in Fig.~\ref{fig:fig3} (but not all particles have the same final radius). Thus, we can calculate the luminosity $L^i_a$ at each radius following Eq.~(\ref{eq:luma}), i.e.,
\begin{equation}
\label{eq:luma2}
L_a^i\approx \epsilon^i_a V_i \approx \epsilon^i_a\times 4\pi r_i^2\, \lambda_i.
\end{equation}

\begin{figure}[htbp]
\centering
\includegraphics[width=\hsize,clip]{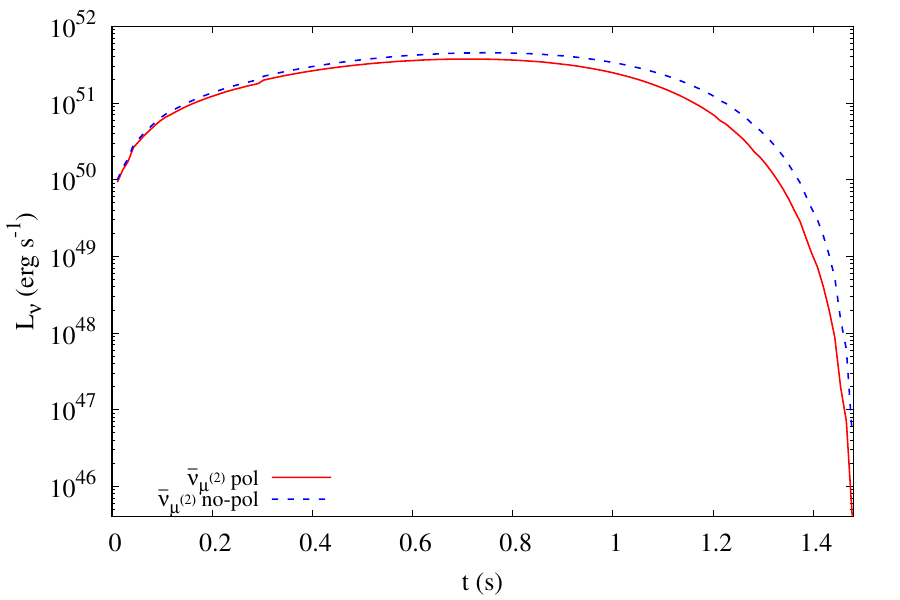}
\caption{$\bar{\nu}_{\mu^{(2)}}$ luminosity, with and without polarization, as a function of time, in the entire time interval of the emission.}
\label{fig:fig11}
\end{figure}

Figure~\ref{fig:fig11} shows the luminosity $L^i_a$ as a function of time, for $a=\bar{\nu}_{\mu^{(2)}}$, within and without considering polarization effects. The luminosity is nonzero only in the region of the ejecta where $pp$ interactions lead to a nonzero production of secondary pions. Therefore, the emission occurs until the shell reaches the radius $r_f=r_4\approx 4.79\times 10^{10}$~cm, after which the proton energy is below the process energy threshold. The emission time is short, from a fraction of a second to $\sim 1$~s.

We can now estimate the total energy emitted in each particle type via Eq.~(\ref{eq:enparticle}). The time interval of the emission, $\Delta t_i$, is the time the shell spends to cover the distance between $r_{i-1}$ and $r_i$ with velocity $\beta_i$, i.e., $\Delta t_i\approx \Delta r_i/(c\,\beta_i)$. Therefore, we have
\begin{equation}
\label{eq:etotintegV}
{\cal E}_a\approx \sum_{i=2}^{n} L^i_a \times \Delta t_i=\sum_{i=2}^n L_a^i\times\frac{\Delta r_i}{c \beta_i},
\end{equation}
where $L_a^i$ is given by Eq.~(\ref{eq:luma2}). The total energy emitted in every particle type, in the whole emitting region, is summarized in Table~\ref{tab:1}.

%%%%%%%%%%%%%%%%%%%%%%%%%%%%%%%%%%%%%%%%%%%%%%%%%%%%%%%%
\subsection{TeV protons interacting with the ISM}\label{sec:4b}
%%%%%%%%%%%%%%%%%%%%%%%%%%%%%%%%%%%%%%%%%%%%%%%%%%%%%%%%

We now consider the interaction of incident protons engulfed by plasma with target protons along the direction of low baryon load $B<10^{-2}$, i.e., with protons of the ISM. Thus, we set the number density of targets as $n_{\rm ISM}\sim 1$~cm$^{-3}$. In this case, the plasma reaches transparency far from the BH site, with an ultrarelativistic Lorentz factor of up to $\gamma_p = \Gamma \sim 10^3$. Therefore, we can assume that incident protons have energies $\sim 1$ TeV. We consider that the interaction occurs in a spherical shell at distances $10^{16}\leq r\leq 10^{17}$~cm from the BH site (see, e.g., \cite{2012A&A...543A..10I}).

Since in this case incident protons have energies $E_p\gtrsim 1$ TeV, we cannot use the parameterization of~\citet{blattnig2000parametrizations} for the cross-section of inelastic $pp$ interaction because its validity is limited to the energy range $(0.3$--$50)$~GeV. Therefore, we follow the approach of~\cite{Kelner2006} to determine the interaction cross-section and spectra of the emerging particles. We recall that~\cite{Kelner2006} studied the $pp$ interactions using the \texttt{SIBYLL}~\cite{PhysRevD.50.5710} and \texttt{QGSJET}~\cite{KALMYKOV199717} codes. For the mentioned energies, we must focus only on their analytical parametrization for the spectra of secondary particles emerging from $\pi$ and $\mu$ decay, and the analytic formula for the energy distribution of pions (for fixed proton energy), for proton energies $E_p\geq 0.1$ TeV and $x=E_a/E_p\geq 10^{-3}$ (where $E_a$ is the energy of the secondary product). Thus, the secondary particles production rate is given by 
\begin{eqnarray}
\label{20}
\Phi_a(E_a)=c\,n_p\int_{E_a}^{\infty} \sigma_{\rm inel}^{pp}(E_p) J_p(E_p) F_a\left(x,E_p\right)\frac{dE_p}{E_p},\quad
\end{eqnarray}
where $n_p$ is the density of target protons, $\sigma_{\rm inel}^{pp}(E_p)$ is the inelastic $pp$ cross-section, and $F_a$ is the specific spectrum for the particle $a$, for which we use the one derived in~\cite{Kelner2006} with an accuracy better than $10\%$.

The inelastic part of the total $pp$ cross-section has been calculated in \cite{Kelner2006} and the results have been shown to be well-fitted by the polynomial
\begin{equation}
\label{21}
\sigma_{\rm inel}^{pp}(E_p)=34.3+1.88\,L+0.25\,L^2\,{\rm mb},
\end{equation}
where $L=\ln(E_p/1\,{\rm TeV})$. This expression for the cross-section is valid for protons energy $E_p>0.1$~TeV. For $E_p\leq 0.1$~TeV, Eq.~\eqref{21} has to be multiplied by the factor $\left[1-(E_p^{\rm Th}/E_p)^4\right]^2$ to take into account the threshold for the pion production, $E_p^{\rm Th}= m_p + 2m_{\pi} + m_{\pi}^2/(2 m_p)$. The parametrization considers $\pi^+$ and $\pi^-$ from $pp$ interactions without distinguishing between electrons and positrons; or neutrinos and antineutrinos. The reason is that the number of $\pi^+$ is larger than the one of $\pi^-$ only by a small amount, and its effect is smaller than the accuracy of the approximations applied in the analysis. Therefore, our calculations also include the contribution of antiparticles (e.g.~$\pi^+$ and $\pi^-$, $\mu^+$ and $\mu^-$).

To get the emissivity of each specific particle from Eq.~(\ref{20}), we must specify the proton energy distribution $J_p\left(E_p\right)$. We consider only protons with fixed energy, so we can write it as $J_p\left(E_p\right)=A \delta\left(E_p-E_p^0\right)$, where $E_p^0$ is our proton fixed energy ($E_p^0=1$ TeV). The constant $A$ is the number density of interacting protons in the considered volume, i.e., $A=N_p/V$. The volume is calculated as $V=(4/3)\pi(r_2^3-r_1^3)$, with $r_1=10^{16}$~cm and $r_2=10^{17}$~cm, and the number of protons is readily obtained from baryon load parameter, i.e., $N_p \approx B E_{e^+e^-}/(m_p c^2)$.

\begin{figure}
\centering
\includegraphics[width=\hsize,clip]{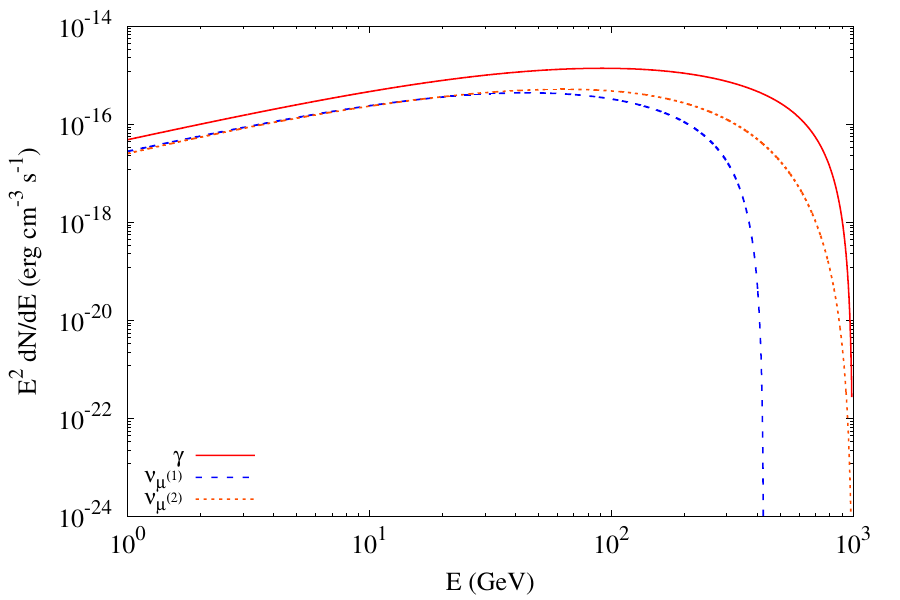}
\caption{Emissivity of high-energy photons (red), neutrino from direct pion decay (blue) and muon neutrino from muon decay (orange) created by the interaction of $E_p=1$~TeV protons with ISM protons at rest ($n_p^{\rm ISM}=1$~cm$^{-3}$).}
\label{LowDensitySpectra}
\end{figure}

In this case, we estimate the luminosity as the one given by the last interaction of the accelerated protons with the target protons of the ISM, namely the ISM shell between $r_{*}=10^{17}-\delta s$~cm and $r_2=10^{17}$~cm, where $\delta s=3\times 10^{10}$~cm is the distance traveled by the protons in $1$~s. This luminosity is given by $L_{a}=\epsilon_a\times \Delta V_{\rm last~shell}$, with $\epsilon_a$ the emissivity of the particle $a$ calculated by Eq.~(\ref{12}) and $\Delta V_{\rm last~shell}=(4/3)\pi(r_2^3-r_{*}^3)$. The total emitted energy for each particle $a$ can be calculated as
\begin{equation}
\label{22}
{\cal E}_{\rm tot,~a}=\Delta t \sum_i \epsilon_{a}^i \times \Delta V_{i},
\end{equation}
where $\Delta V_i=(4/3)\,\pi \left(r_i^3-r_{i-1}^3\right)$, with $r_{i}=r_{i-1}+\delta s$, and $\Delta t$ the plasma crossing time of the ISM shell. Since $\epsilon_{a}^i$ does not depend on the radius, the total emitted energy can be obtained by
\begin{equation}
\label{23}
{\cal E}_{\rm tot,~a}=\epsilon_a~\Delta t ~\Delta V,
\end{equation}
with $\Delta V=4/3\,\pi \left(r_2^3-r_1^3\right)$. The photon emissivity from $\pi^0$ decay is given by Eq.~(\ref{20}), with $F_{\gamma}\left(x,E_p\right)$ derived by \cite{Kelner2006}. We note that their parameterization of the photon spectrum includes the photons produced by the different decay channels of $\eta$ mesons. Figure~\ref{LowDensitySpectra} shows the photon emissivity. The total energy emitted through photons in all emitting regions via Eq.~(\ref{23}) is ${\cal E}_{\gamma}=5.41\times 10^{43}$~erg. The luminosity emitted in the last emitting shell of the ISM region, calculated as explained above, is $L_{\gamma}=1.01\times 10^{43}$~erg~s$^{-1}$. The photon spectrum peaks at $E_{\gamma}=91.62$~GeV.

The muon neutrino from direct pion decay ($\pi\rightarrow \mu \nu_{\mu}$) is given by the same Eq.~(\ref{20}), with $F_{\nu_{\mu^{(1)}}}\left(E_{\nu_{\mu}}/E_p^0,E_p\right)$ derived in \cite{Kelner2006}. The emissivity is shown in Fig.~\ref{LowDensitySpectra}. The figure shows that the spectrum has a sharp cut-off at $x=0.427$. This effect is due to the kinematics of the process since, at high energy, this neutrino can take only a factor $\lambda=1-r_\pi=0.427$ of the pion energy. The total energy emitted in $\nu_{\mu^{(1)}}$ inside the whole emitting region is ${\cal E}_{\nu_{\mu^{(1)}}}= 1.60\times 10^{43}$~erg, the luminosity in the last emitting shell is $L_{\nu_{\mu^{(1)}}}=3.01\times 10^{42}$~erg~s$^{-1}$, and the spectrum peaks at the neutrino energy $E_{\nu_{\mu^{(1)}}}=44.72$~GeV.

The muon neutrino luminosity from muon decay can be calculated from Eq.~(\ref{20}), with the specific $F_a\left(x, E_p\right)$ for each particle derived in \cite{Kelner2006}. The spectrum of $\nu_{\mu^{(2)}}$ and $\nu_e$ can be represented by the same function (with an error less than $5\%$ for $\nu_e$). The emissivity is shown in Fig.~\ref{LowDensitySpectra}. Differently from the $\nu_{\mu^{(1)}}$, the energy of these particles can reach, at most, the energy of the $\mu$, which is not limited to taking a specific amount of the pion energy ($E_{\mu}^{\rm max}\approx E_{\pi}$). The total energy emitted in $\nu_{\mu^{(2)}}$ in the emitting region is ${\cal E}_{\nu_{\mu^{(2)}}}=1.98\times 10^{43}$~erg, the luminosity of the last emitting shell is $L_{\nu_{\mu^{(2)}}}=3.71\times 10^{42}$~erg~s$^{-1}$, and the spectrum peaks at $E_{\nu_{\mu^{(2)}}}=63.9$~GeV.

%%%%%%%%%%%%%%%%%%%%%%%%%%%%%%%%%%%%%%%%%%%%%%%%%%%%%%%%
%%%%%%%%%%%%%%%%%%%%%%%%%%%%%%%%%%%%%%%%%%%%%%%%%%%%%%%%
\section{Summary, Discussion and Conclusions}\label{sec:5}
%%%%%%%%%%%%%%%%%%%%%%%%%%%%%%%%%%%%%%%%%%%%%%%%%%%%%%%%
%%%%%%%%%%%%%%%%%%%%%%%%%%%%%%%%%%%%%%%%%%%%%%%%%%%%%%%%

In this article, we have addressed the main features of possible production channels of neutrinos in long GRBs within the BdHN model. We have shown that these systems can produce neutrinos of MeV and GeV-TeV energies by different mechanisms. 

In Section~\ref{sec:3}, we summarized the role of neutrino flavor oscillations in two astrophysical systems in the BdHN
model of GRBs: spherical accretion onto an NS and disk accretion around the newborn BH from the collapsed NS companion of the CO star. In both systems, the emission of large amounts of neutrinos mediates the system's cooling, allowing the accretion process to reach rates between ($10^{-4}$--$10^{-2}$)~$M_\odot$~s$^{-1}$. The ambient conditions of density and temperature in both physical situations imply a high flux of neutrinos of energies of up to a few tens of MeV (see Table \ref{tab:tab1} and Fig. \ref{fig:Disks}). Furthermore, those ambient conditions lead to flavor oscillations dictated by self-interaction potentials and the MSW resonance.

In the case of spherical accretion, the neutrino emission is dominated by the annihilation of $e^+e^-$ pairs, leading to a $0.7$--$0.3\%$ distribution between electron and non-electron neutrino flavors, respectively. The hierarchy between the oscillation potentials shown in Fig.~\ref{fig:potentials} implies that neutrino self-interactions dominate the flavor evolution close to the neutron star surface, leading to bipolar oscillations with oscillation lengths between $50$ m and $1$ km. The symmetry between neutrinos and antineutrinos induces kinematic decoherence and neutrino flavor equipartition. When the new flavor distribution leaves the accretion region, the matter potential suppresses further oscillations until the MSW resonance occurs. Thus, the neutrino flavor distribution arising from the accretion region differs from the flavor distribution at the NS surface. Specifically, Table~\ref{tab:tabfluxes} shows that the distribution becomes $\sim$ $0.55$--$0.45$ for normal hierarchy and $\sim$ $0.62$--$0.38$ for inverted hierarchy.

In the case of disk accretion onto the newborn Kerr BH, we built a simple NCAD disk model that accounts for neutrino oscillations within it. We showed that the oscillation dynamic affects the disk's cooling and flavor emission. As in spherical accretion, Figs.~\ref{fig:Disks} and \ref{fig:Potentials1} show that the system produces neutrinos and antineutrinos in comparable numbers. The potentials follow a particular hierarchy in which self-interactions dominate the flavor evolution with fast oscillations. Extending the results in~\cite{Raffelt:2007yz,2007JCAP...12..010F,EstebanPretel:2007ec} to our system, we conclude that flavor equipartition is the final state of the accretion disk. Therefore, the flavor content emitted by the disk differs from the one in traditional models~\cite{2012PhRvD..86h5015M, 2016PhRvD..93l3004L}.

The bottom panels of Fig.~\ref{fig:Potentials1} summarize the differences between standard disks and the ones with flavor equipartition. The latter increases the cooling efficiency by allowing electron neutrinos to escape as non-electron neutrinos. Equation~(\ref{eq:fluxcomp}) gives the ratio between the neutrino cooling flux for a disk with and without flavor equipartition. The surfeit of non-electron neutrinos reduces the disk's sensibility to the opacity of electron neutrinos and reduces the energy density of the $e^{-}e^{+}$ plasma around the BH.

In general, strong gravitational fields influence neutrino flavor oscillations. We have taken into account general relativistic effects in the determination of the dynamics of the accretion process and the behavior of the thermodynamic variables in the spacetime around the central object (NS or BH), but not directly in the equations governing the neutrino flavor oscillations. Setting up the general relativistic equations of neutrino self-interactions is a non-trivial problem that goes beyond the scope of our paper. Indeed, the literature on this subject is limited, and most calculations deal with the simplified problem of single neutrino trajectories in a vacuum (assuming the neutrino as a massless test particle) or, at best, in a constant electron background, but not for neutrino ensembles. Those prescriptions of the problem account for the phase shift between different mass eigenstates which depends on the neutrino 4-momentum (see, e.g., \cite{1996GReGr..28.1161A, PhysRevD.55.7960, PhysRevD.56.1895, PhysRevD.101.024016}).

For the disk accretion, our approximation has been to consider that the neutrino oscillations occur in a sufficiently small vicinity around each disk point such that gravity effects are negligible in such a small region. Hence, we adopt the neutrino oscillation framework in flat spacetime. The neutrino flavor oscillation length is smaller than one percent of the characteristic size of the disk, which validates our assumption. Therefore, we expect the inclusion of gravitational effects in the neutrino oscillation equations to affect our conclusions negligibly.

For the spherical accretion around the NS, within the lightbulb approximation, the gravitational redshift affects the neutrino energy and the vacuum potential. The gravitational bending affects the neutrino trajectories, hence the self-interaction potential. Only highly symmetric spacetimes would allow performing the trajectory average analytically. As for the case of disk accretion, we expect the above additional gravitational effects to produce minor quantitative changes but not the hierarchy of the potentials. Therefore, we expect the qualitative picture and our conclusions to hold.

In Section \ref{sec:4}, we have computed the neutrino production via $pp$ interactions in BdHN I. The accretion process and the BH formation lead to density asymmetries in the SN ejecta. Therefore, the $e^+e^-$ plasma created in the BH formation process engulfs different amounts of matter of the surrounding SN ejecta during its expansion and self-acceleration depending on the direction (see Fig.~\ref{fig:Interactionsscheme1}). This asymmetry leads to a direction-dependent Lorentz factor for the engulfed protons in the expanding shell.

From this scheme, we have studied two types of physical setups for $pp$ interactions that cover the generality of the system. In Section~\ref{sec:4a}, we studied the $pp$ interactions in the high-density regime, i.e., inside the SN ejecta. We have performed relativistic hydrodynamical simulations of the $e^+e^-$ plasma for fixed baryon load parameter~(see, e.g., \cite{2018ApJ...852...53R}), using the PLUTO code~\cite{2019ApJS..242...20M,mignone2011pluto}. We use the baryon load parameter $B=51.75$ as an example, for which the protons engulfed by the plasma acquire energies up to $\sim 7$ GeV ($\gamma_p\leq 7$) due to the plasma Lorentz factor. The number of target protons is $\sim 10^{23}$~cm$^{-3}$, while the number density of incident protons is $\sim 10^{25}$~cm$^{-3}$.

The obtained spectra show that the neutrinos and photons have energies $E_{\nu_{\mu^{(1)}}}\leq 2$~GeV and $E_{\gamma},~E_{\nu_{\mu^{(2)}}},~E_{\nu_e}< 5$~GeV.
Particle production occurs in the first $\sim 1.5$~s of the shell expansion (see Fig.~\ref{fig:fig11}). After, the proton energy falls below the threshold for $pp$ interaction with target protons in the remnant. The luminosity emitted in $\nu$ is $L_{\nu}=(2$--$3)\times 10^{51}$~erg~s$^{-1}$ and, since the emission occurs in~$1.5$~s, the associated total integrated energy is $\sim 10^{50}$--$10^{51}$~erg (see Table~\ref{tab:1} for the total energy of each particle). Summing up the total energy released in all the secondary $\nu$, we obtain that they correspond to $3\%$ of the initial plasma energy, $E_{e^{\pm}}$, and protons carry off $14\%$ of it.

In Section~\ref{sec:4b}, we have considered the expansion of the $e^+e^-$ plasma in the direction of low baryon load, where we adopted $B=10^{-3}$~(see, e.g., \cite{ruffini2016classification,2018ApJ...852...53R}). The expanding $\gamma e^{\pm}$ plasma engulfs baryons in the \textit{cavity} around the BH~\cite{2019ApJ...883..191R}. The self-acceleration brings the engulfed protons to energies of up to $E_p\sim 1$~TeV ($\gamma_p\sim 10^3$). In this case, we obtained the wider range of particles energies, $1\leq E_{a}\leq 10^3$~GeV, with an associated total luminosity of $L_{\gamma}=1.0135\times 10^{43}$~erg~s$^{-1}$,~$L_{\nu_{\mu^{(1)}}}=3.01\times 10^{42}$~erg~s$^{-1}$, and $L_{\nu_{\mu^{(2)}},\nu_e}=3.71\times 10^{42}$~erg~s$^{-1}$.

A precise estimate of the detection probability of these neutrinos lies outside the scope of this present work. However, based on the present results, we can estimate the detection probability of these neutrinos for some of Earth's neutrino detectors. We focus our attention on three detectors: SuperKamiokande \cite{fukuda2003super}, HyperKamiokande \cite{yokoyama2017hyper}, and IceCube \cite{abbasi2012design}. The two Kamiokande detectors explore a wide energy range for neutrinos (from a few MeV up to $100$~PeV). The IceCube detector works principally on high-energy neutrinos ($\gtrsim$ PeV), but the core of the experiment (the \textit{Deep Core Detector}) works until energies of the order of $10$~GeV. For our estimation, we need only the effective area of the detector, which we denote by $A_{\rm eff}$.

The number of neutrinos per unit of area that arrive at the detector can be estimated as
\begin{equation}
\label{24}
\frac{dN_{\nu}}{dS}=\frac{{\cal E}_{\nu}}{4\pi D^2 E^*_{\nu}},
\end{equation}
where ${\cal E}_{\nu}$ is the total energy emitted in neutrino (see Table~\ref{tab:1}), $D$ is the luminosity distance to the source, $E_{\nu}$ is the neutrino energy. For the latter, we use the value at the end of the specific particle spectrum for the high-density case and at the spectrum's peak in the low-density case. $E^*_{\nu}=E_{\nu}/(1+z)$ is the redshifted neutrino energy, and $z$ the source cosmological redshift. Therefore, we can obtain the number of detectable neutrinos as $N^{\rm det}_{\nu}=A_{\rm eff} \times dN_\nu/dS$.

Using the approximate equation for the luminosity distance, $c\,z \approx H_0 D$, valid for relatively low cosmological redshift, where $H_0$ is the Hubble constant (we use $H_0=72$~km~s$^{-1}$~Mpc$^{-1}$), we can obtain the neutrino-detection horizon
\begin{equation}
\label{26}
D_h=\frac{K H_0}{2c}+\frac{1}{2}\sqrt{\frac{K^2 H_0^2}{c^2}+4K},
\end{equation}
which is the luminosity distance for which $N^{\rm det}_{\nu}=1$. Here, $K={\cal E}_{\nu}A_{\rm eff}/(4\pi E_\nu)$. Table~\ref{tab:Dh} summarizes $D_h$ for $\nu_{\mu^{(1)}}$ and $\nu_{\mu^{(2)}}$, in the high and low-density regions, for the three considered detectors. 

\begin{table}[t]
\centering
\caption{Horizon distances $D_h$ for $\nu_{\mu}$'s from direct pion decay and $\mu$ decay, for the high and low-density region cases, for the three considered detectors: SuperKamiokande (SK), HyperKamiokande (HK), IceCube-DeepCore (Deep). The effective area of the detectors have been taken from~\cite{abe2021search} (for SK),~\cite{abe2011letter,abe2018hyper} (for HK) and~\cite{abbasi2012design} (for DeepCore). For the effective area of DeepCore, we use the open square line in Fig. 8 of~\cite{abbasi2012design}. For SK, we considered the effective area for the up-going muon averaged over the zenith angle. Note that we consider the IceCube detector only for the higher energy regime since the effective area of the detector arrives at a minimum of $10$~GeV (see~\cite{abbasi2012design}). For the high-density region, we considered the neutrinos from the $\pi^+$ decay $\left(\nu_{\mu^{(1)}},\bar{\nu}_{\mu^{(2)}}\right)$ since they have the highest total energy (see Table~\ref{tab:1}).}
\begin{tabular}{*{6}{c}}
\hline
\hline
    Particle  &  $D^{\rm SK}_h$  & $D^{\rm HK}_h$ & $D^{\rm Deep}_h$ & $E_{\nu}$ & ${\cal E}_{\nu}$ \\%& $A_{\rm eff}$  \\
    \hline
\textbf{High density}  &&&&&  \\
\textbf{region}    &   (kpc)  & (Mpc) & - - & (GeV)    & ($10^{51}$~erg)  \\% & ( cm$^2$)  \\
\hline
${\nu_{\mu^{(1)}}}$ & $43.68$ & $2.20$ & - - & $2.03$ & $0.47$ \\%& $1.58\times 10^{-6}$ \\
\hline
$\nu_{\mu^{(2)}}$ & $516.12$ & $7.37$ & - - & $4.75$ & $3.53$ \\%& $6.87\times 10^{-5}$ \\
\hline
\textbf{Low density} &&&&&  \\
   \textbf{region}  & (pc)  & (kpc) & (kpc) & (GeV) & ($10^{43}$~erg)\\%& (cm$^2$)\\
\hline
$\nu_{{\mu}^{(1)}}$  & $136.88$ & $1.22$ & $2.77$ & $44.72$  & $1.60$ \\ %& \\
\hline
$\nu_{\mu^{(2)}}$ & $180.35$ & $1.56$ & $3.38$ & $63.9$  & $1.98$ \\%& \\
\hline
\hline
\end{tabular}
\label{tab:Dh}
\end{table}

The values of $D_h$ in Table \ref{tab:Dh} imply that it is unlike the detection of GeV-TeV neutrinos from BdHN I emitted by the production channels evaluated in this work. The probability of occurrence of a BdHN I at such close distance is extremely low (see, e.g., \cite{2016ApJ...832..136R, 2018ApJ...859...30R}). We advance the possibility that BdHN I might produce very high-energy neutrinos. The inner engine of the high-energy emission in BdHN I might produce those neutrinos along (or close to) the rotation axis of the BH, where it accelerates electrons to energies of up to $10^{18}$~eV (and protons up to $10^{21}$~eV) \cite{2019ApJ...886...82R, rueda2020blackholic, 2021A&A...649A..75M, 2021MNRAS.504.5301R, 2022ApJ...929...56R}. This topic remains of interest for future research.

Detecting the photons emitted in these $pp$ interactions might indirectly test the present physical picture of neutrino emission. The photons arising from the interactions in the low-density region have energies of hundreds of GeV and a low luminosity $\sim 10^{43}$~erg~s$^{-1}$ (see Section~\ref{sec:4b}). For a source at $z=1$ ($D\approx 6.7$~Gpc), the above leads to a photon flux on Earth of $\sim 10^{-15}$~erg~s$^{-1}$~cm$^{-2}$, which is below the flux threshold of the Large Area Telescope (LAT) of the \textit{Fermi} satellite (see \cite{ajello2019decade}, for the second GRB catalog of \text{Fermi}-LAT).

The photons from the high-density region have energies of the order of a few GeV and a luminosity of $10^{51}$-$10^{52}$~erg~s$^{-1}$ (see, e.g., Table~\ref{tab:1}). For a source at the same distance considered above, these photons have a flux at Earth of $\sim 10^{-7}$--$10^{-6}$~erg~s$^{-1}$~cm$^{-2}$, a value sufficiently high to be detected by Fermi-LAT. 
The photons from the high-density region have energies of the order of a few GeV and a luminosity of $10^{51}$-$10^{52}$~erg~s$^{-1}$ (see, e.g., Table~\ref{tab:1}). For a source at the same distance considered above, these photons have a flux at Earth of $\sim 10^{-7}$--$10^{-6}$~erg~s$^{-1}$~cm$^{-2}$, a value sufficiently high to be detected by Fermi-LAT. However, the emission radii of $\sim 10^{10}$~cm, together with the aforementioned high photon luminosity, lead to high opacity and, consequently, to the impossibility of these photons leaving the system. In Appendix~\ref{app:1}, we calculate the photon's mean free path for the main photon opacity mechanisms that we expect in these systems. Unfortunately, the results imply that the photons created inside the ejecta cannot freely escape the system due to the high opacity.

\appendix

%%%%%%%%%%%%%%%%%%%%%%%%%%%%%%%%%%%%%%%%%%%%%%%
\section{Photon opacity}
\label{app:1}
%%%%%%%%%%%%%%%%%%%%%%%%%%%%%%%%%%%%%%%%%%%%%%%
%

There are several photon interaction mechanisms leading to their opacity:~1) \textsl{photo-meson production}: $\gamma+p\rightarrow h+~n\pi$ (where $h$ is a hadron and $n$ the number of produced pions);~2) \textsl{photon-proton pair-production}: $\gamma+p\rightarrow p~e^+~e^-$ (Bethe-Heitler process);~3) \textsl{Compton scattering}: $\gamma+p\rightarrow \gamma^{'}~p$ (where $\gamma^{'}$ is the photon emerging with different energy in comparison with to the interacting one);~4) \textsl{pairs Compton scattering}: $\gamma+~e^{\pm}\rightarrow \gamma^{'}+~e^{\pm}$;~5) \textsl{photon pair production}: $\gamma+\gamma\rightarrow e^+ e^-$ (Breit-Wheeler process). 

We calculate the mean free path, $l_{\rm int}=\left(\sigma_{\rm int}~n_{\rm targ}\right)^{-1}$, for the main processes. Here, $\sigma_{\rm int}$ is the cross-section for the considered interaction, and $n_{\rm targ}$ is the number density of targets. We consider here as the main processes the Breit-Wheeler and the Bethe-Heitler process.

%%%%%%%%%%%%%%%%%%%%%%%%%%%%%%%%%%%%%%%%%%%%%%%%%%%%%%%%
\subsection{Photon-Proton pair-production}
\label{Behte-Heitler}
%%%%%%%%%%%%%%%%%%%%%%%%%%%%%%%%%%%%%%%%%%%%%%%%%%%%%%%%
%

In this calculation, we assume a photon energy $E_{\gamma}=0.69$~GeV, corresponding to the peak of the higher photon spectrum (see, e.g., the curve corresponding to the radius $r_1$ in~Fig.~\ref{fig:fig5}). The Bethe-Heitler pair production results as the dominant process for $E_{\gamma}>10$~MeV. At $E_{\gamma}>500$~MeV, the probability to have pair production via this process is close to unity; see, e.g., Fig.~(34.17) in \cite{Zyla:2020zbs} where this probability is calculated for various absorbing elements. For infinite photon energy, the cross-section is
\begin{equation}
\label{ppp4}
\sigma_{pp}(\infty)= \frac{7}{9}\left(\frac{A}{X_0 N_A}\right),
\end{equation}
with $X_0$ is the interaction length, $A$ the element molar mass and $N_A$ the Avogadro's number (see Ref.~\cite{Zyla:2020zbs}). At a finite photon energy $E_{\gamma}$, the cross-section becomes \cite{tsai1974pair}
\begin{equation}
    \label{pp5}
    \sigma_{pp}\left(E_{\gamma}\right)=\sigma_{pp}(\infty) \left(1-\xi\right),
\end{equation}
with $\xi=\left[\sigma_{pp}(\infty)-\sigma_{pp}\left(E_{\gamma}\right)\right]/\sigma_{pp}(\infty)$. Considering hydrogen ($A=1$~g/mol,~$Z=1$), the value of $\xi$ (for a photon energy of $E_{\gamma}\simeq 0.7$~GeV) is $\xi =0.174$ (see Ref.~\cite{tsai1974pair}). Thus, the cross-section becomes $\sigma_{pp}\simeq 17.07$~mb. For a number density of targets $\langle n \rangle\sim 6.5\times 10^{23}$~cm$^{-3}$ (see Section~\ref{S:PhysicalQuantities}), we obtain $l_{\rm int}\simeq 90.13$~cm.

%%%%%%%%%%%%%%%%%%%%%%%%%%%%%%%%%%%%%%%%%%%%%%%%%%%%%%%%
\subsection{Photon-Photon Pair Production}
\label{pairprod-gamma-gamma}
%%%%%%%%%%%%%%%%%%%%%%%%%%%%%%%%%%%%%%%%%%%%%%%%%%%%%%%%

In this estimation, we calculate the mean free path at each step of the plasma expansion inside the ejecta. We consider only the photons produced by interactions inside the SN ejecta. The mean free path at the radius $i$ is given by $l^i_{\gamma\gamma}=\left(\sigma^{i}_{\gamma\gamma} n^i_{\gamma}\right)^{-1},$ with $n^i_{\gamma}$ the photon number density, and $\sigma^{i}_{\gamma\gamma}$ the pair production cross-section \cite{gould1967pair}
\begin{equation}
\label{app1}
    \sigma_{\gamma\gamma}=\frac{3}{16}(1 - \zeta^2)\left[2\zeta
(\zeta^2 - 2) + (3 - \zeta^4)\ln\left(\frac{1 + \zeta}{1 -
\zeta}\right)\right]\sigma_{\rm T},
\end{equation}
being $\sigma_{\rm T}=(8 \pi/3)e^4/(m_e c^2)^2=6.65\times10^{-25}$~cm$^2$ the Thomson cross-section, $\zeta=\sqrt{1-2
(m_e c^2)^2/[\epsilon_1\,\epsilon_2 \left(1-\cos\theta\right)]}$ the velocity of the particle in the center of the momentum frame, and $\theta$ is the interaction angle between two photons.
We assume an isotropic radiation field since, at every radius, the photons are produced with the same energy by the same mechanism. As for the photon energy, at each radius, we consider the energy at the spectrum peak (see Fig.~\ref{fig:fig5}).

\begin{figure}[h]
\centering
\includegraphics[width=\hsize,clip]{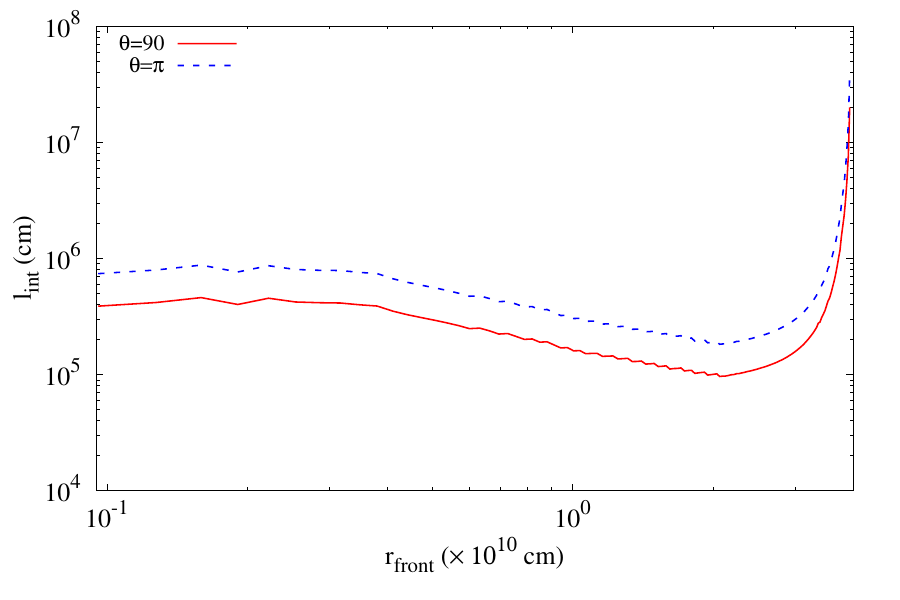}
\caption{Interaction length for $\gamma \gamma$ pair production, for photons produced and interacting inside the ejecta, for two angles of interaction between the photons, $\theta=\pi/2$ (red) and $\pi$ (dashed-blue).}
\label{fig:l_int_gammagamma}
\end{figure}

We estimate the photons number density by $n^i_{\gamma}=L^i_{\gamma}/(4\pi r_i^2 c \varepsilon^i_{\gamma})$, where $L^i_{\gamma}$ is the photon luminosity, calculated at every radius. The resulting mean free path is shown in Fig.~\ref{fig:l_int_gammagamma}, for two interaction angles $\theta= \pi$ (head-on collision) and $\theta=\pi/2$. Therefore, we obtain a mean free path in the range $l^i_{\gamma\gamma}\sim 10^5$--$10^6$ cm.

%%%%%%%%%%%%%%%%%%%%%%%%%%%%%%%%%%%%%%
% \begin{acknowledgments}
% ...
% \end{acknowledgments}

\bibliographystyle{apsrev4-2}
\bibliography{references,references_Stefano}

\end{document}